\documentclass[floatfix,eqsecnum,preprintnumbers,aps,nofootinbib]{revtex4} 

\usepackage{epsfig}
\usepackage{citesort}

\def \bsllg     {$B_s \to \ell^+ \ell^- \gamma$ }

\def \btosllg   {$b \to s \ell^+ \ell^- \gamma$ }
\def \btosll    {$b \to s \ell^+ \ell^-$ }
\def \btoxsga   {$B \to X_s \gamma$ }

\def \cq#1      {C_{Q_#1} }
\def \c#1eff    {C_#1^{eff} }

\def \pl        {{\rm $P_L$ }}
\def \pn        {{\rm $P_N$ }}
\def \pt        {{\rm $P_T$ }}
 
\def \epsmualbesig  {\varepsilon_{\mu\alpha\beta\sigma} }
\def \modmsq    {|{\cal M}|^2 }
\def \modm#1sq  {|{\cal M}_#1|^2 }
\def \rem12     {Re({\cal M}_1 {\cal M}^*_2) }
\def \facmatrix {| \frac{\alpha^{3/2} G_F}{\sqrt{2 \pi}} V_{tb}
                 V_{ts}^*|^2 } 
\def \modsq#1   {|#1|^2}
\def \rea#1#2   {Re(#1^* #2)}
\def \facdr     {| \frac{\alpha^{3/2} G_F}{2 \sqrt{2 \pi}} V_{tb}
                 V_{ts}^*|^2 } 
\def \ctencqtwo {\left(C_{10} + \frac{m_{B_s}^2}{2 m_\ell m_b}
C_{Q_2}\right)} 
\def \drcqone   {\left(\frac{m_{B_s}^2}{2 m_\ell m_b} C_{Q_1}\right)}
\def \lnzh      {{\rm ln}(\hat{z})}
\def \dgdsdcost {\frac{d^2 \Gamma}{d \sh d \cos\theta}}

\def \mle#1     {m_\ell^#1 }
\def \mbs       {m_{B_s}}
\def \mb#1      {m_{B_s}^#1}
\def \ppp        {p_+}
\def \ppm        {p_-}

\def \psq       {p^2}
\def \p#1qsq    {(p_#1 q)^2}
\def \pqsq      {(p q)^2}
\def \ppqsq     {(p_+ q)^2}
\def \pmqsq     {(p_- q)^2}

\def \sh        {\hat{s}}
\def \mlh       {\hat{m}_\ell}
\def \mlhsq     {\hat{m}_\ell^2}
\def \zh        {\hat{z}}
\def \faco      {\sqrt{1 - \frac{4 \hat{m}_\ell^2}{\hat{s}}}}

\def \pmv       {{\bf p}_-}
\def \ppv       {{\bf p}_+}
\def \qv        {{\bf q}}
\def \ev#1      {{\bf e}_#1}
\def \wv#1      {{\bf w}_#1}

\def \beq       {\begin{equation}}
\def \eeq       {\end{equation}}
\def \beqa      {\begin{eqnarray}}
\def \eeqa      {\end{eqnarray}}
\def \bfig      {\begin{figure}}
\def \efig      {\end{figure}}
\def \bcen      {\begin{center}}
\def \ecen      {\end{center}}

\def \ie        {{\it i.e. }}

\def \etal      {{\it et. al. }}

\def \prd#1#2#3       {Phys. \ Rev. {\bf D #1}, {#2} (#3)}
\def \pr#1#2#3        {Phys. \ Rev. {\bf #1}, {#2} (#3)}
\def \prl#1#2#3       {Phys. \ Rev. \ Lett. {\bf #1}, {#2} (#3)}
\def \nuclphysb#1#2#3 {Nucl. \ Phys. {\bf B #1}, {#2} (#3)}
\def \plb#1#2#3       {Phys. \ Lett. {\bf B #1}, {#2} (#3)}
\def \physrep#1#2#3   {Phys. \ Rep {\bf #1}, {#2} (#3)}
\def \zphysc#1#2#3    {Z. \ Phys. {\bf C #1}, {#2} (#3)}

\begin{document}

\preprint{hep-ph/0205nnn}

\title{Supersymmetric effects in \bsllg decays} 

\author{S. Rai Choudhury}
 \email{src@ducos.ernet.in}
\author{Naveen Gaur}
 \email{naveen@physics.du.ac.in} 

\affiliation{ Department of Physics and Astrophysics, \\
     University of Delhi, \\
     Delhi - 110 007, India.}

\date{\today}


\begin{abstract}
B-meson decays are very useful probes for testing the Standard Model
and its various extensions. Leptonic decays of B have very clean
signatures in this respect and hence can be very useful testing
grounds. In this work we study the effects of MSSM (Minimal
Supersymmetric Extension of Standard Model) on various kinematical
distributions in the radiative dileptonic decay (\bsllg). We study the
Forward Backward asymmetry (of the lepton pair), and the various
polarization asymmetries of both final state leptons ($\ell^-$ and
$\ell^+$). In radiative dileptonic decay of B-meson (\bsllg ) the
final state photon can also be polarized. So in this channel one can
also study the polarization effects of the final state photon.  
\end{abstract}

\maketitle




\section{\label{section:1}Introduction}

There are many theoretical and experimental reasons for studying
flavor-changing neutral current (FCNC) processes. These transitions
are forbidden at tree level in Standard Model (SM) and hence provide
 very stringent tests of SM at loop level. The investigations of
various FCNC processes can  be used to accurately determine various
fundamental parameters of SM like elements of CKM matrix, various
decay constants etc. Besides testing SM the FCNC processes can be very
useful for discovering indirect effects of possible TeV scale
extensions of SM like SUSY (Supersymmetry). In particular, the
processes like \btoxsga , $B \to X_s \ell^+ \ell^-$, \bsllg etc. are
experimentally very clean and possibly are very sensitive to various
extensions of SM. Compared to \btoxsga, the flavor-changing
semi-leptonic decays (like $B \to X_s \ell^+ \ell^-$, $B \to K^*
\ell^+ \ell^-$, \bsllg etc.) can be more sensitive to the actual form
of new interactions since here one can measure experimentally various
kinematical distributions apart from total decay rate. The various
kinematical distributions which can be measured in above mentioned
processes can be Forward Backward asymmetry, CP asymmetry , various
lepton polarization asymmetries etc. 

\par The flavor changing channel \btosll decay, which takes place in
SM at loop level is very sensitive to the gauge structure of SM. This
 mode (\btosll) is also very sensitive to the various extensions of
SM. New physics (here we are concerned with SUSY) effects manifests in
rare B decays in essentially two different ways, either through the
new contributions to the existing Wilson coefficients in SM or through
the new structures in effective Hamiltonian which were absent in
SM. Particularly \btosll has been extensively studied within SM and in
various extensions of it
\cite{Ali:1997vt,Skiba:1993mg,Dai:1997vg,Choudhury:1999ze,Xiong:2001up,Huang:1999vb,RaiChoudhury:1999qb,goto1,Bobeth:2001jm,Cho:1996we,Hewett:1996ct,Grossman:1997qj,Kruger:1996cv,Grinstein:1989me,Long-Distance}.
Final state lepton polarizations (there can be in general three 
polarizations namely longitudinal, normal and transverse) provides a
very useful probe for establishing new physics
\cite{Fukae:1999ww,RaiChoudhury:1999qb,RaiChoudhury:2002i}. 

\par The simplest and the most favorite extension of the SM has been
the Minimal Supersymmetric Standard Model (MSSM). As we know that
there are five physical scalars (Higgs) in MSSM as compared to one in
SM. The importance of specially Neutral Higgs Bosons (NHBs) \btosll
has been extensively discussed earlier
\cite{Iltan:2000iw,Skiba:1993mg,Dai:1997vg,Choudhury:1999ze,Xiong:2001up,Huang:1999vb,RaiChoudhury:1999qb,Bobeth:2001jm,Grossman:1997qj,RaiChoudhury:2002i}.
The importance of lepton polarization in \bsllg has already been
pointed out in earlier work \cite{RaiChoudhury:2002i}. In this work we
present the complete study of all the lepton polarization asymmetries
associated with final state leptons in radiative dileptonic decay
(\bsllg) within the framework of Minimal Supersymmetric Standard Model
(MSSM) . Here we will do the combined analysis the various
polarization asymmetries associated with both final state leptons
($\ell^-$ and $\ell^+$) within MSSM. In the radiative dileptonic decay
mode (\bsllg ) apart from lepton polarization we can also have
polarized photon (\ie photon of a particular helicity). We will also
study the effects of polarized photon by introducing a variable {\sl
``photon polarization asymmetry'' }. This is perhaps a very difficult
parameter to be measured experimentally but we include it for
completeness. With this variable we now have one more kinematical
variable (along with total decay rate, FB asymmetry and three
different lepton polarization asymmetries ) to test out the exact
structure of effective Hamiltonian.  In all our analysis we will try
to focus on mainly the NHB effects.

\par This paper is organized as follows : In section \ref{section:2}
we will first present the QCD corrected effective Hamiltonian of the
quark level process \btosllg, including NHB effects. We will give the
corresponding matrix element and then will give the analytic
expression of the dilepton invariant mass distribution and forward
backward asymmetry. In section
\ref{section:3} we will give the analytical expressions of the various
lepton polarization asymmetries. In section \ref{section:4} we analyse
the combinations of the polarization asymmetries. Then in section
\ref{section:4b} we will move to another kinematical variable {\sl
photon polarization asymmetry}. We will finally conclude in section
\ref{section:5} with results.  

\section{\label{section:2} Dilepton invariant mass distribution}

The exclusive \bsllg decay is induced by the inclusive \btosll
one. So, we have to start with QCD corrected effective Hamiltonian for
related quark level process \btosll, which can be obtained by
integrating out heavy particles in SM and MSSM \cite{Choudhury:1999ze,Huang:1999vb,Bobeth:2001jm}
\beq
{\cal H} ~=~ - ~\frac{4 G_F}{\sqrt{2}} V_{tb} V_{ts}^* 
       \sum_{i=1}^{10} \Bigg\{ C_i(\mu) O_i(\mu) ~+~ 
      C_{Q_i}(\mu) Q_i(\mu)
      \Bigg\}
\label{sec1:eq:1}
\eeq
where $O_i$ are current-current ($i = 1,2$), penguin ($i =
1,\dots,6$), magnetic penguin ($i = 7,8$) and semileptonic ($i =
9,10$) operators. $Q_i , (i = 1,\dots,10)$ are the operators which
results due to NHB exchange diagrams
\cite{Huang:1999vb,Bobeth:2001jm}. $C_i(\mu) $ and $C_{Q_i}(\mu)$ are
Wilson coefficients evaluated at scale $\mu$ and are tabulated in
\cite{Choudhury:1999ze, Huang:1999vb,Bobeth:2001jm,Xiong:2001up}.

\par Neglecting the strange quark mass the effective Hamiltonian
(\ref{sec1:eq:1}) give the following matrix element for the inclusive
\btosll :
\beqa
{\cal M} &=& \frac{\alpha G_F}{2 \sqrt{2} \pi} V_{tb}V_{ts}^*
             \left\{ ~ -2 ~ \c7eff ~\frac{m_b}{p^2} ~
                 \bar{s} i\sigma_{\mu\nu}p^\nu (1+\gamma_5) b
	             ~ \bar{\ell} \gamma^\mu \ell 
                  ~+~ \c9eff ~\bar{s}\gamma_\mu(1-\gamma_5) b
                   ~\bar{\ell}\gamma^\mu\ell  \right.    \nonumber \\
         &&   \left. ~+ ~C_{10} ~\bar{s}\gamma_\mu (1-\gamma_5) b
                      ~\bar{\ell}\gamma^\mu\gamma_5\ell
                    ~+~ \cq1 ~\bar{s} (1+\gamma_5) b
                        ~\bar{\ell}\ell
                    ~+~ \cq2 ~\bar{s}(1+\gamma_5)b
                        ~\bar{\ell}\gamma_5\ell ~
              \right\}
\label{sec1:eq:2}
\eeqa
where $p ~=~ \ppp + \ppm$ is the sum of momenta of $\ell^-$ and
$\ell^+$, $V_{tb}, V_{ts}$ are CKM factors. The coefficients
$C_7^{eff}, C_9^{eff}, C_{10}$ and $C_{Q_1}, C_{Q_2}$ are given in
many earlier works
\cite{Choudhury:1999ze,Huang:1999vb,Bobeth:2001jm}. We 
will also take the long distance effects related to the charm
resonances according to the Briet Wigner form
\cite{Xiong:2001up,RaiChoudhury:1999qb,Kruger:1996cv,Long-Distance}

\par In order to obtain the matrix element for \bsllg decay, a photon
line should be hooked to any of the charged internal or external
lines. As has been pointed out before \cite{Aliev:1997ud},
contributions coming from hooking a photon line from any charged
internal line will be suppressed by a factor of $m_b/M_W^2$, hence we
neglect them in our further analysis. When photon is attached to the
initial quark lines the corresponding matrix element is the {\sl so 
called} {\bf  structure dependent} (SD) part of the amplitude which
can be written as :
\beqa
{\cal M}_{SD} = \frac{\alpha^{3/2} G_F}{\sqrt{2 \pi}} V_{tb} V_{ts}^* 
           &&  \left\{ ~[ A ~\epsmualbesig {\epsilon^*}^\alpha p^\beta
                  q^\sigma ~+~ i B ~
                 (\epsilon_\mu^*(pq)-(\epsilon^*p)q_\mu)] ~ 
                 \bar{\ell}\gamma^\mu\ell
               \right.                                 \nonumber\\
           && + ~ \left. [ C ~\epsmualbesig {\epsilon^*}^\alpha p^\beta
                   q^\sigma ~+~ i D ~
                   (\epsilon_\mu^*(pq)-(\epsilon^*p)q_\mu) ] ~
                   \bar{\ell}\gamma^\mu\gamma_5\ell   ~
                \right\}
\label{sec1:eq:3}
\eeqa
where 
\beqa
A  &=&  \frac{1}{m_{B_s}^2} ~[ \c9eff G_1(p^2) ~-~  2 \c7eff
        \frac{m_b}{p^2}G_2(p^2)],                   \nonumber  \\ 
B  &=&  \frac{1}{m_{B_s}^2}~ [ \c9eff F_1(p^2) ~-~ 2 \c7eff
        \frac{m_b}{p^2}F_2(p^2)],                   \nonumber  \\ 
C  &=&  \frac{C_{10}}{m_{B_s}^2}~G_1(p^2),           \nonumber  \\
D  &=&  \frac{C_{10}}{m_{B_s}^2}~F_1(p^2).
\label{sec1:eq:4}
\eeqa
In getting eqn.(\ref{sec1:eq:3}) we have used following definitions of
the form factors \cite{Eilam:1995zv} 
\beqa
\langle\gamma |~ \bar{s} \gamma_\mu (1 \pm \gamma_5) b ~|B_s \rangle
   &=&   \frac{e}{m_{B_s}^2}
         \left\{ \epsmualbesig \epsilon_\alpha^* p_\beta q_\sigma 
             G_1(p^2)\mp i [ (\epsilon_\mu^*(pq)-(\epsilon^*p)q_\mu) ]
            F_1(p^2) 
         \right\}                                          
\label{sec1:eq:5}       \\
\langle\gamma| ~\bar{s} i \sigma_{\mu\nu} p_\nu (1 \pm \gamma_5) b ~
 |B_s\rangle 
  &=&   \frac{e}{m_{B_s}^2}
        \left\{  \epsmualbesig \epsilon_\alpha^* p_\beta q_\sigma 
           G_2(p^2) \pm i [ (\epsilon_\mu^*(pq)-(\epsilon^*p)q_\mu) ]
           F_2(p^2) 
        \right\}
\label{sec1:eq:6}     
\eeqa
another relation we can get by multiplying $p_\mu$ on both the sides
of eqn.(\ref{sec1:eq:6} :
\beq
\langle\gamma|~ \bar{s} (1 \pm \gamma_5) b ~|B_s\rangle ~=~ 0
\label{sec1:eq:7}
\eeq
Here $\epsilon_\mu$ and $q_\mu$ are the four vector polarization and
momentum of photon respectively. We can see from eqn.(\ref{sec1:eq:2})
that neutral scalar exchange parts  doesn't contribute to the {\bf
structure dependent} part. 

\par When the photon is attached to the lepton lines using the
expressions :
\beqa
\langle 0|~\bar{s} b ~|B_s\rangle &=& 0 
\label{sec1:eq:8}           \\
\langle 0|~\bar{s}\sigma_{\mu\nu} (1 + \gamma_5) b~|B_s\rangle &=& 0 
\label{sec1:eq:9}                       \\
\langle 0|~ \bar{s} \gamma_\mu \gamma_5 b ~|B_s\rangle &=& -~ i
f_{B_s} P_{B_s\mu}
\label{sec1:eq:10}
\eeqa
and the conservation of vector current we can get the contribution to
the Bremsstrahlung part (called {\bf internal Bremsstrahlung} IB) part
as :
\beqa
{\cal M}_{IB}
   &=&  \frac{\alpha^{3/2} G_F}{\sqrt{2\pi}} ~V_{tb} V_{ts}^* ~
       i 2 ~m_{\ell} ~
        f_{B_s} ~
        \left\{ ~( C_{10} ~+~ \frac{m_{B_s}^2}{2 m_{\ell} m_b} \cq2) ~
           \bar{\ell} \left[ \frac{\not\epsilon \not P_{B_s}}{2 \ppp q}
              ~-~ \frac{\not P_{B_s}\not\epsilon}{2 \ppm q}
                      \right] \gamma_5\ell
        \right.                              \nonumber        \\
   &&  ~+~ \left. \frac{m_{B_s}^2}{2 m_{\ell} m_b} \cq1 
           \left[ 2 m_\ell ( \frac{1}{2 \ppm q} ~+~ \frac{1}{2 \ppp q})~ 
                  \bar{\ell}\not\epsilon\ell
        ~+~ \bar{\ell} ~ ( \frac{\not\epsilon \not P_{B_s}}{2 \ppp q}
        ~-~ \frac{\not P_{B_s}\not\epsilon}{2 \ppm q}) ~ \ell ~
          \right] ~
        \right\}.
\label{sec1:eq:11}
\eeqa
where $P_{B_s}$ and $f_{B_s}$ are the momentum and decay constant of
the $B_s$ meson. $\ppm$ and $\ppp$ are the four momental of $\ell^-$ and
$\ell^+$ respectively. 

\par The total matrix element for \bsllg is obtained as a sum of
${\cal M}_{SD}$ and ${\cal M}_{IB}$ terms :
\beq
{\cal M} ~=~ {\cal M}_{SD} ~+~ {\cal M}_{IB}
\label{sec1:eq:12}
\eeq
From above matrix element we can get the square of the
matrix element as,(with photon polarizations summed over) 
\beq
\sum_{\rm photon ~pol} \modmsq ~=~ |{\cal M}_{SD}|^2 ~+~ |{\cal M}_{IB}|^2
~+~ 2 Re({\cal M}_{SD} {\cal M}_{IB}^*) 
\label{sec1:eq:13}
\eeq
with 
\beqa
|{\cal M}_{SD}|^2 &=& 4 ~ \facmatrix ~ 
             \left\{ 
            ~[ ~\modsq{A} + \modsq{B}~ ] ~ [ \psq ( \pmqsq + \ppqsq ) 
               + 2 m_\ell^2 \pqsq ] ~+~ [~ \modsq{C} + \modsq{D}~ ] 
	     \right.                  \nonumber   \\
         &&  \left. ~ [ \psq ( \pmqsq +  \ppqsq ) - 2 m_\ell^2 \pqsq ]
                ~+~ 2 ~Re(B^* C + A^*D)~ \psq ( \ppqsq - \pmqsq ) 
             \right\}    
\label{sec1:eq:14}             \\
|{\cal M}_{IB}|^2  &=& 
   4 ~ \facmatrix ~ f_b^2 ~ \mle2 ~
       \Bigg[
	  \ctencqtwo 
           \left\{ ~ 8 + 
              \frac{1}{\pmqsq } ( - 2 \mb2 \mle2 - \mb2 \psq + p^4 ~+~
                 2 p^2 ( \ppp q ) )
           \right.                    \nonumber  \\
 &&        \left.  
         .  ~+~  \frac{1}{(\ppm q)} ( 6 \psq + 4 (\ppp q) )
         ~+~  \frac{1}{\ppqsq } 
           ( - 2 \mb2 \mle2 - \mb2 \psq + p^4 + 2 \psq (\ppm q) )
         ~+~  \frac{1}{(\ppp q)} ( 6 \psq + 4 (\ppm q) )
           \right.                      \nonumber   \\
 &&        \left. 
         ~+~ \frac{1}{(\ppm q) (\ppp q)} ( - 4 \mb2 \mle2 + 2 p^4 )
            ~ \right\}                  \nonumber    \\
 &&      +~  \drcqone 
         \left\{ 8  ~+~  \frac{1}{\pmqsq } 
         . ( 6 \mb2 \mle2 + 8 \mle4 - \mb2 \psq - 8 \mle2 \psq 
            + p^4 - 8 \mle2 (\ppp q) + 2 \psq (\ppp q) ) 
         \right.                      \nonumber    \\
 &&      \left. 
             +~ \frac{1}{ (\ppm q) } (- 40 \mle2 + 6 \psq + 4 (\ppp q) )
             ~+~  \frac{1}{\ppqsq } ( 6 \mb2 \mle2 + 8 \mle4 - \mb2
           \psq - 8 \mle2 \psq + p^4 - 8 \mle2 (\ppm q) 
         \right.                      \nonumber   \\
 &&      \left.  +~ 2 \psq (\ppm q) )
             ~+~ \frac{1}{ (\ppp q) } ( - 40 \mle2 + 6
              \psq  + 4 (\ppp q) )  
         \right.  \nonumber   \\
 &&      \left. 
        +~ \frac{1}{(\ppm q) (\ppp q)} ( 4 \mb2 \mle2 + 16 \mle4
                 - 16 \mle2 p^2  + 2 p^4 ) 
         \right\}
	\Bigg]                                    
\label{sec1:eq:15}              
\eeqa
\beqa
2 Re( {\cal M}_{SD} {\cal M}_{IB}^* ) 
 &=&  
16 ~ \facmatrix ~ f_b ~ \mle2 
       \Bigg[  \ctencqtwo 
               \left\{ ~ - ~ Re(A) \frac{( \ppm q + \ppp q)^3}{(\ppm q)
                      (\ppp q)}
               \right.               \nonumber    \\
 &&            \left. +~ Re(D) \frac{ (p q)^2 ( \ppm q ~-~ \ppp q ) }
                        {(\ppm q) (\ppp q)} 
               \right\} 
               ~+~ \drcqone ~ 
               \left\{ Re(B) ~ \frac{1}{(\ppm q)(\ppp q)} ( - (p q)^3
               \right.                 \nonumber    \\ 
 &&            \left.  - 2 (\ppm \ppp) (\ppp q)^2 
                 - 2 (\ppm \ppp) (\ppm q)^2 
                + 4 \mle2 (\ppm q)(\ppp q)) 
                +~ Re(C) \frac{(p q)^2 ( \ppm q - \ppp q )}{(\ppm q)
                   (\ppp q)} 
               \right\}
\label{sec1:eq:16}
\eeqa
The differential decay rate of \bsllg as a function of invariant mass
of lepton pair is given by:
\beqa
\frac{d \Gamma}{d\sh} 
   = ~\facdr ~\frac{\mb5}{16 (2 \pi)^3}~ (1 - \sh) ~ \faco
      ~ \bigtriangleup 
\label{sec1:eq:17}
\eeqa
with $\bigtriangleup$ defined as
\beqa
\bigtriangleup &=& ~~ 
      {4 \over 3} ~\mb2 ~( 1 - \sh)^2~ 
        [ ~ ( \modsq{A} + \modsq{B} )~ (2 \mlhsq + \sh) 
          ~+~ ( \modsq{C} + \modsq{D} ) ( - 4 \mlhsq + \sh) ~ ]
                                                     \nonumber \\
  &&  +~ \frac{64 ~f_b^2 \mlhsq}{\mbs^2} ~\ctencqtwo^2 ~ \frac{[~ (1 - 4 
          \mlhsq + \sh^2) \lnzh ~-~ 2 \sh \faco ~ ] }{(1 -
          \sh)^2 ~ \faco}                            \nonumber \\
  &&  -~ \frac{64 ~f_b^2 \mlhsq}{\mbs^2} ~ \drcqone^2 ~ 
         \frac{[~ ( - 1 + 12 \mlhsq - 16 \mlh^4 - \sh^2)
              \lnzh ~+~ ( -2 \sh - 8 \mlhsq \sh + 4 \sh^2 ) \faco ~]}
              {(1 - \sh)^2 ~ \faco }        \nonumber \\ 
  &&  +~ 32 ~ f_b \mlhsq ~\ctencqtwo ~ Re(A) ~\frac{\lnzh}{\faco} 
                              \nonumber  \\
  &&  
      -~ 32 ~ f_b \mlhsq ~\drcqone ~Re(B)~ \frac{[ ~ (1 - 4
          \mlhsq + \sh) \lnzh ~-~ 2 \sh \faco ~]}{\faco}
\label{sec1:eq:18}
\eeqa
where $\sh = \psq/\mb2 ~,~ \mlhsq = m_\ell^2/\mb2 ~,~ \zh = \frac{1 + 
\faco}{1 - \faco}$ are dimensionless quantities. 

\bfig[ht]
\begin{center}
\epsfig{file=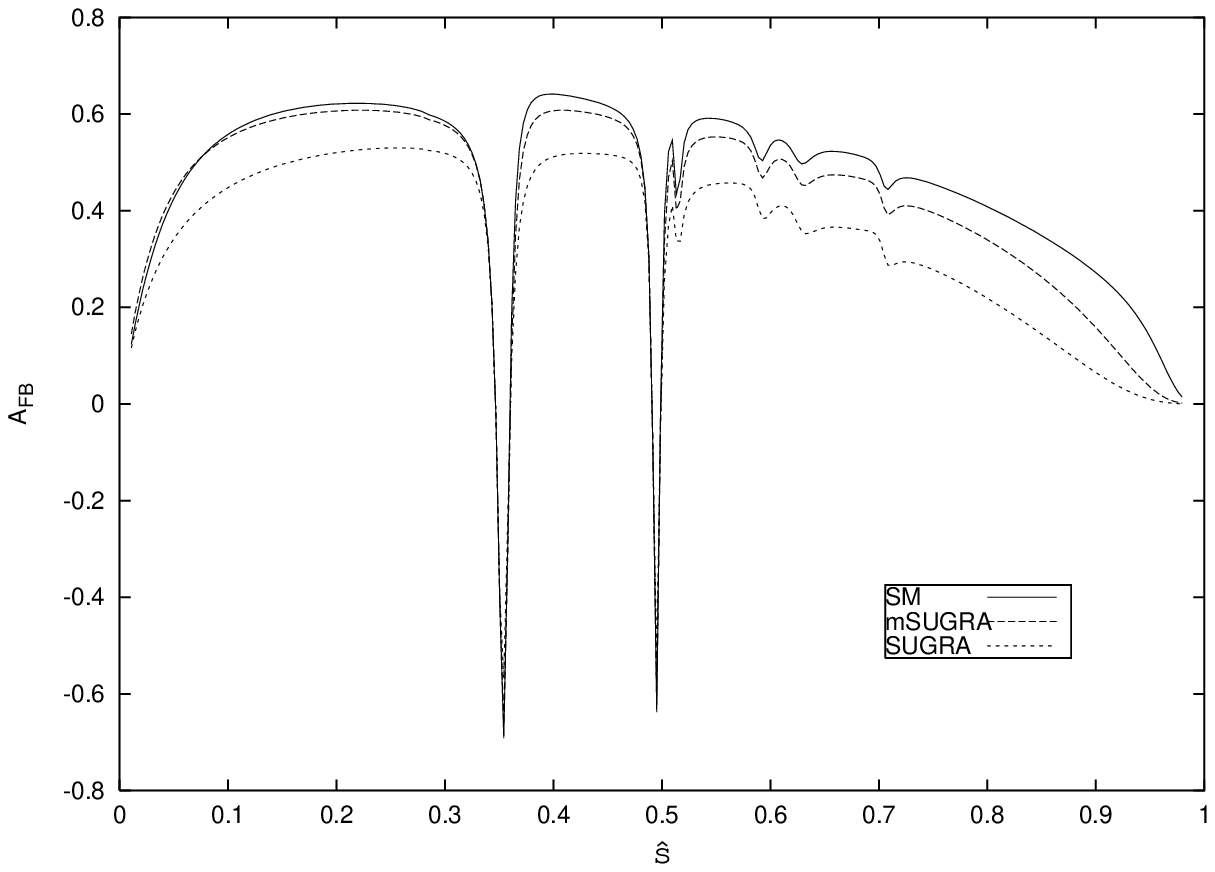,width=3.5in}
\epsfig{file=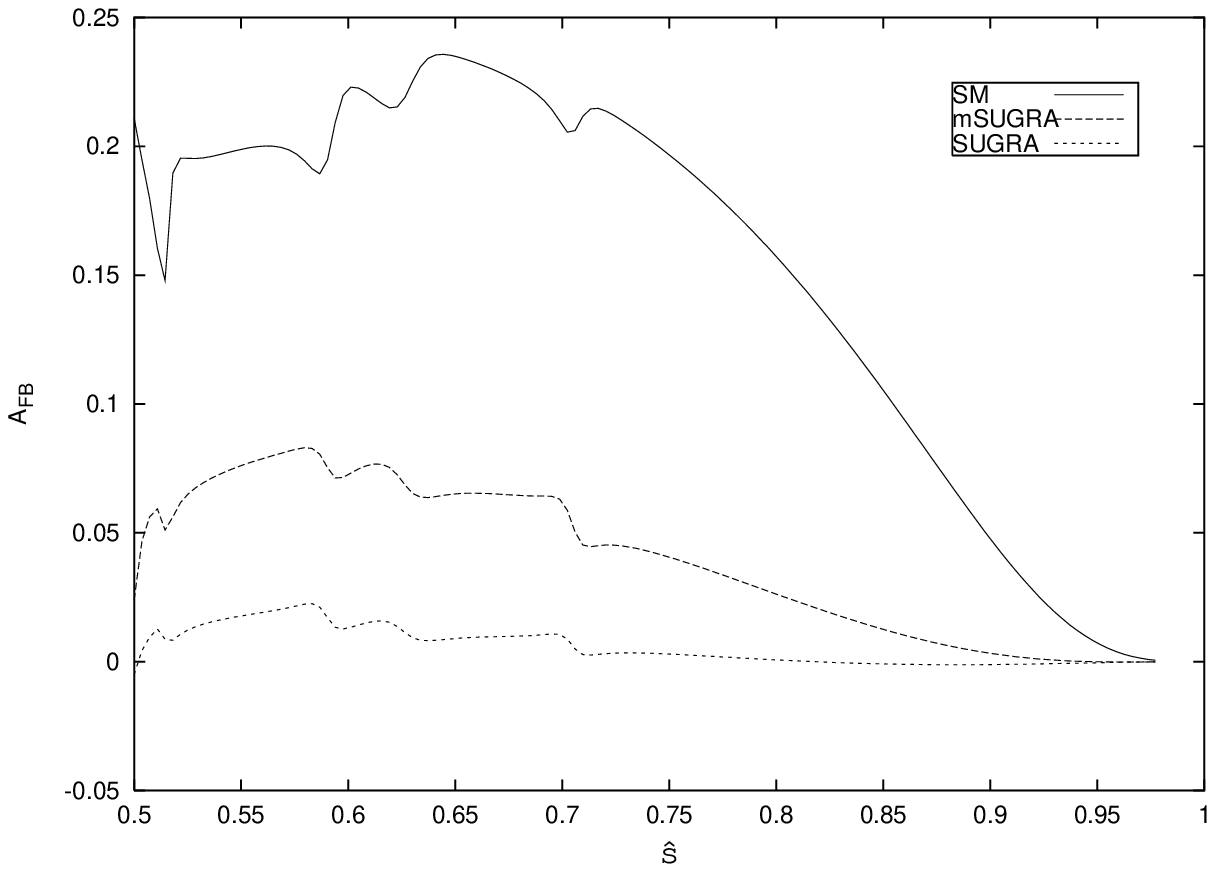,width=3.5in}
\caption{Forward Backward asymmetry of $\mu$ (left) and $\tau$
(right). Parameters are : $m = 200 , ~M = 450 ,~ A = 0 , ~tan\beta =
40$ and sgn($\mu$) is +ve. For SUGRA the pseudoscalar Higgs mass is
taken  to be 306. All masses are in GeV}
\label{fig:1}
\end{center}
\efig

\par The forward-backward (FB) asymmetry is also very sensitive to the
details of the new physics (SUSY here). We can define the FB asymmetry
as :
\beq
A_{FB} ~=~  
   \frac{ \int_0^1 d \cos\theta \dgdsdcost - \int_{-1}^0 d\cos\theta
       \dgdsdcost }{\int_0^1 d \cos\theta \dgdsdcost + \int_{-1}^0
       d\cos\theta \dgdsdcost }
\label{sec1:eq:19}
\eeq
where $\theta$ is the angle between momentum of B-meson and $\ell^-$
in the center of mass frame of dilepton. The analytical expression of
FB asymmetry is :
\beqa
A_{FB} &=& 
      \Bigg[
        - 2 ~\mb2 ~Re(A^* D + B^* C) ~(1 - \sh)^2 ~\sh ~ \faco 
         ~~+~~ 32 ~f_b ~ \mle2 ~ \frac{(-1 + \sh)}{\faco} ~
           Log\left(\frac{4 \hat m_\ell^2 }{\sh} \right) 
                             \nonumber     \\
     && \times \left\{ ~\ctencqtwo ~Re(D) ~+~ \drcqone~ Re(C) ~ \right\} 
      \Bigg]/\bigtriangleup
\label{sec1:eq:20}
\eeqa

\section{\label{section:3} Lepton polarization asymmetries}

We compute noe the polarization asymmetries from the four fermi
interactions defined in eqn.(\ref{sec1:eq:3}) and
eqn.(\ref{sec1:eq:11}). We define the following orthogonal unit
vectors $S$ in the rest frame of $\ell^-$ and $W$ in the rest frame of
$\ell^+$, for the polarizations of the leptons \cite{Kruger:1996cv} to
the longitudinal direction (L), the normal directions (N) and the
transverse direction (T)
\beqa
S^\mu_L &\equiv& (0, \ev{L}) ~=~ (0, \frac{\pmv}{|\pmv|})  
                 \nonumber               \\
S^\mu_N &\equiv& (0, \ev{N}) ~=~ (0, \frac{\qv \times \pmv}{|\qv
             \times \pmv})      
                 \nonumber               \\
S^\mu_T &\equiv& (0, \ev{T}) ~=~ (0, \ev{N} \times \ev{L}) 
            \label{sec2:eq:1}             \\
W^\mu_L &\equiv& (0, \wv{L}) ~=~ (0, \frac{\ppv}{|\ppv|})  
                 \nonumber                \\
W^\mu_N &\equiv& (0, \wv{N}) ~=~ (0, \frac{\qv \times \ppv}{|\qv
             \times \ppv})      
                 \nonumber                \\
W^\mu_T &\equiv& (0, \wv{T}) ~=~ (0, \wv{N} \times \wv{L}) 
                 \label{sec2:eq:2}            
\eeqa
where $\ppv$, $\pmv$ and $\qv$ are three momenta of $\ell^+$, $\ell^-$
and photon respectively in the CM frame of $\ell^- \ell^+$ system. Now
on boosting above vectors defined by
eqns.(\ref{sec2:eq:1},\ref{sec2:eq:2}) to the CM frame of $\ell^-
\ell^+$ system. Only the longitudinal vector will get boosted while
the other two (transverse and normal) will remain the same. The
longitudinal vectors after the boost will become
\beqa
S^\mu_L &=& \left( \frac{|\pmv|}{m_\ell}, \frac{E_1 \pmv}{m_\ell
                  |\pmv|}   \right) 
           \nonumber               \\
W^\mu_L &=& \left( \frac{|\pmv|}{m_\ell}, - \frac{E_1 \pmv}{m_\ell
                |\pmv|} \right) 
\label{sec2:eq:3}
\eeqa
The polarization asymmetries can now be calculated by using the spin
projector ${1 \over 2}(1 + \gamma_5 \not S)$ for $\ell^-$ and the
spin projector is ${1 \over 2}(1 + \gamma_5 \not W)$ for $\ell^+$
. The lepton polarization asymmetries can be defined as :
\beqa
P_x^- &\equiv & \frac{( \frac{d\Gamma( S_x, W_x )}{d\sh} + 
                        \frac{d\Gamma( S_x, - W_x )}{d\sh} )
                       - ( \frac{d\Gamma( - S_x, W_x )}{d\sh}
                         + \frac{d\Gamma( - S_x, - W_x )}{d\sh} )}
                      {( \frac{d\Gamma( S_x, W_x )}{d\sh} + 
                        \frac{d\Gamma( S_x, - W_x )}{d\sh} )
                       + ( \frac{d\Gamma( - S_x, W_x )}{d\sh}
                         + \frac{d\Gamma( - S_x, - W_x )}{d\sh} )}, \\
P_x^+ &\equiv & \frac{( \frac{d\Gamma( S_x, W_x )}{d\sh} + 
                        \frac{d\Gamma( - S_x, W_x )}{d\sh} )
                       - ( \frac{d\Gamma( S_x, - W_x )}{d\sh}
                         + \frac{d\Gamma( - S_x, - W_x )}{d\sh} )}
                      {( \frac{d\Gamma( S_x, W_x )}{d\sh} + 
                        \frac{d\Gamma( S_x, - W_x )}{d\sh} )
                       + ( \frac{d\Gamma( - S_x, W_x )}{d\sh}
                         + \frac{d\Gamma( - S_x, - W_x )}{d\sh} )}
\label{sec2:eq:4}
\eeqa
where the sub-index x is L,T or N. \pl denotes the longitudinal
polarization asymmetry, \pt is the polarization asymmetry in the
decay plane and \pn is the normal component to both of them. \pl and
\pt are P-odd, T-even and hence CP even observable. But \pn is P-even,
T-odd and hence a CP-odd observable \footnote{because time reversal
operation changes the sign of the momentum and spin, and parity
transformation changes only the sign of momentum}.
 
\bfig[h]
\begin{center}
\epsfig{file=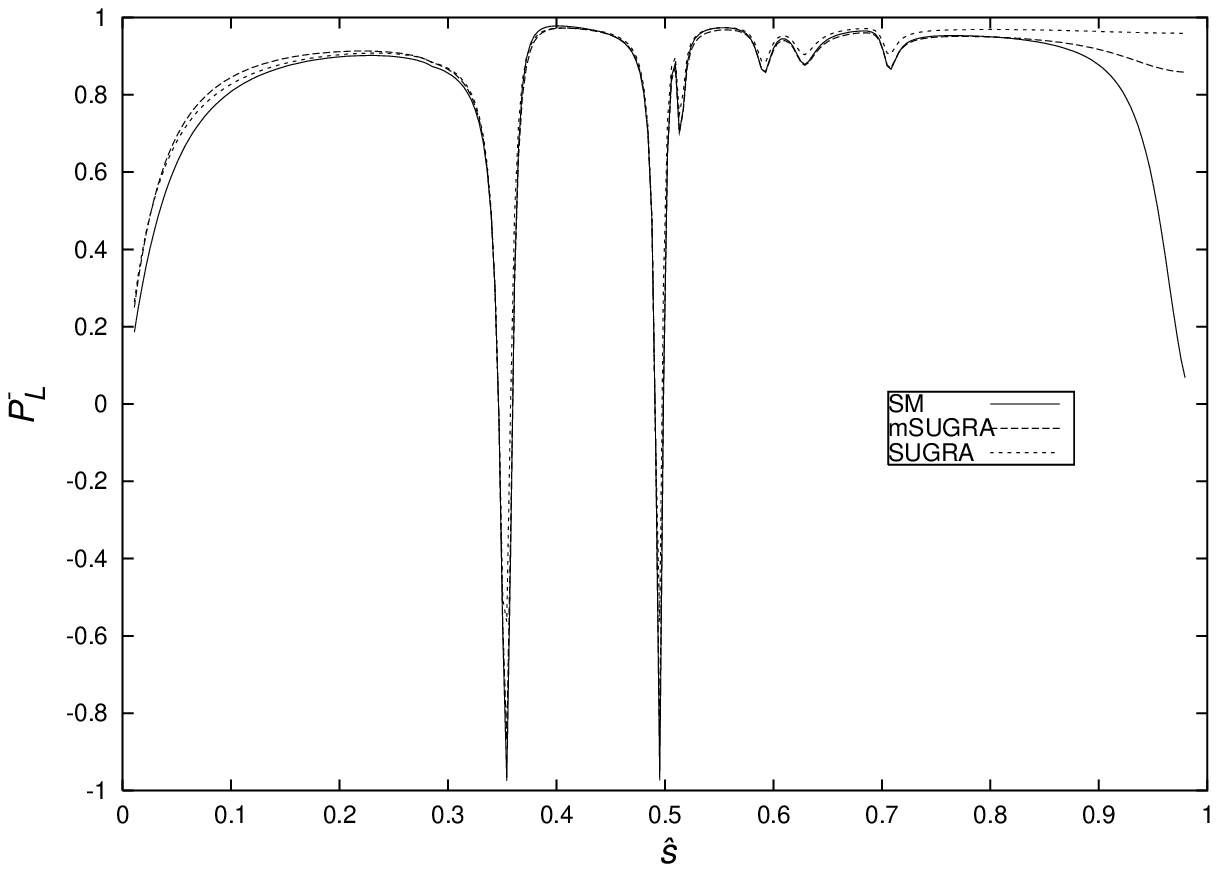,width=3.5in}
\epsfig{file=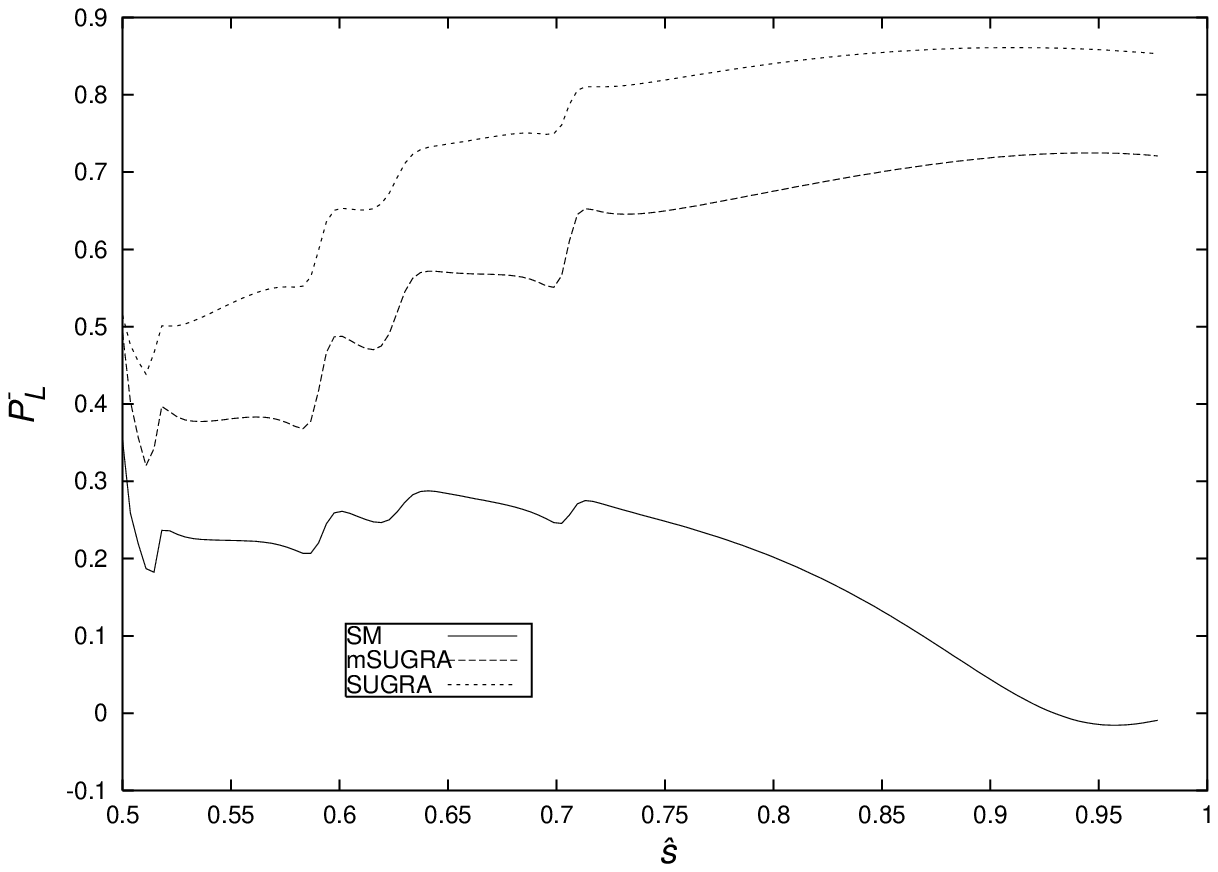,width=3.5in}
\caption{Longitudinal Polarization asymmetry of $\ell ^+ \mu$ (left)
and $\tau$  (right). Parameters are : $m = 200 , ~M = 450 ,~ A = 0 ,
~tan\beta = 40$ and sgn($\mu$) is +ve. For SUGRA the pseudoscalar
Higgs mass is taken to be 306. All masses are in GeV}
\label{fig:2}
\end{center}
\efig

\par The longitudinal polarization asymmetries for the leptons are :
\beqa
P_L^- &=& 
  \Bigg[
     \frac{8 \mb2 }{3} ~Re(~A^* C ~+~ B^* D~) ~L_1(\sh) 
   +~  128 ~\frac{f_b^2 \mle2 }{\mb2 } \ctencqtwo \drcqone L_2(\sh) 
             \nonumber   \\
 && +~ 32 \ctencqtwo f_b \mle2 ~\left\{~ Re(B) L_3(\sh) ~+~ Re(C)
         L_4(\sh) ~ \right\}       
             \nonumber   \\ 
 && +~ 32 \drcqone f_b ~\mle2 ~\left\{ ~ Re(A) L_5(\sh) ~+~ Re(D)
          L_6(\sh) ~ \right\}
  \Bigg]/\bigtriangleup
\label{sec2:eq:5}                    \\
P_L^+ &=& 
  \Bigg[
    - ~ \frac{8 \mb2 }{3}~ Re(~ A^* C ~+~ B^* D ~) ~L_1(\sh) 
   +~  128 ~\frac{f_b^2 \mle2 }{\mb2 } \ctencqtwo \drcqone L_2(\sh) 
             \nonumber   \\
 && +~ 32 \ctencqtwo f_b ~\mle2 ~\left\{~ Re(B) L_3(\sh) ~-~ Re(C)
         L_4(\sh) ~\right\}       
             \nonumber   \\ 
 && +~ 32 \drcqone f_b ~\mle2 ~\left\{ ~ Re(A) L_5(\sh) ~-~ Re(D)
        L_6(\sh) ~ \right\}
  \Bigg]/\bigtriangleup
\label{sec2:eq:6}
\eeqa
with
\beqa
L_1(\sh) &=& \sh (1 - \sh)^2 \faco      \nonumber  \\
L_2(\sh) &=& \frac{\left\{( \sh - 4 \mle2 \sh - 2 \sh^2 - 4 \mle2
                 \sh^2 + 3 \sh^3 ) \faco  ~+~ ( 2 \mle2 - 8 \mle4 -
                 \sh + 8 \mle2 \sh - 8 \mle4 \sh + 2 \mle2 \sh^2 -
                 \sh^3 ) ln(\zh)  \right\}}{(1 - \sh)^2 ( \sh - 4
                 \mle2 )}        \nonumber  \\
L_3(\sh) &=& \frac{4 \mle2 - \sh - 12 \mle2 \sh + 3 \sh^2 + ( 2 \mle2
                + 2 \mle2 \sh - 2 \sh^2) ln(\zh) \faco}{(4 \mle2 -
                  \sh) \faco} 
                                 \nonumber \\
L_4(\sh) &=& \frac{(-1 + \sh) \left\{ \sh \faco + (2 \mle2 - \sh) ln(\sh)
             \right\} }{(4 \mle2 - \sh)}  \nonumber  \\
L_5(\sh) &=& \frac{ (-1 + \sh) ( - \sh \faco + 2 \mle2 ln(\zh) ) }{(\sh
             - 4 \mle2 )}        \nonumber \\
L_6(\sh) &=& \frac{ (-1 + \sh) ( \sh \faco ~+~ (2 \mle2 - \sh )
                ln(\zh) }{4 \mle2 - \sh } 
\label{sec2:eq:7}
\eeqa

\bfig[h]
\begin{center}
\epsfig{file=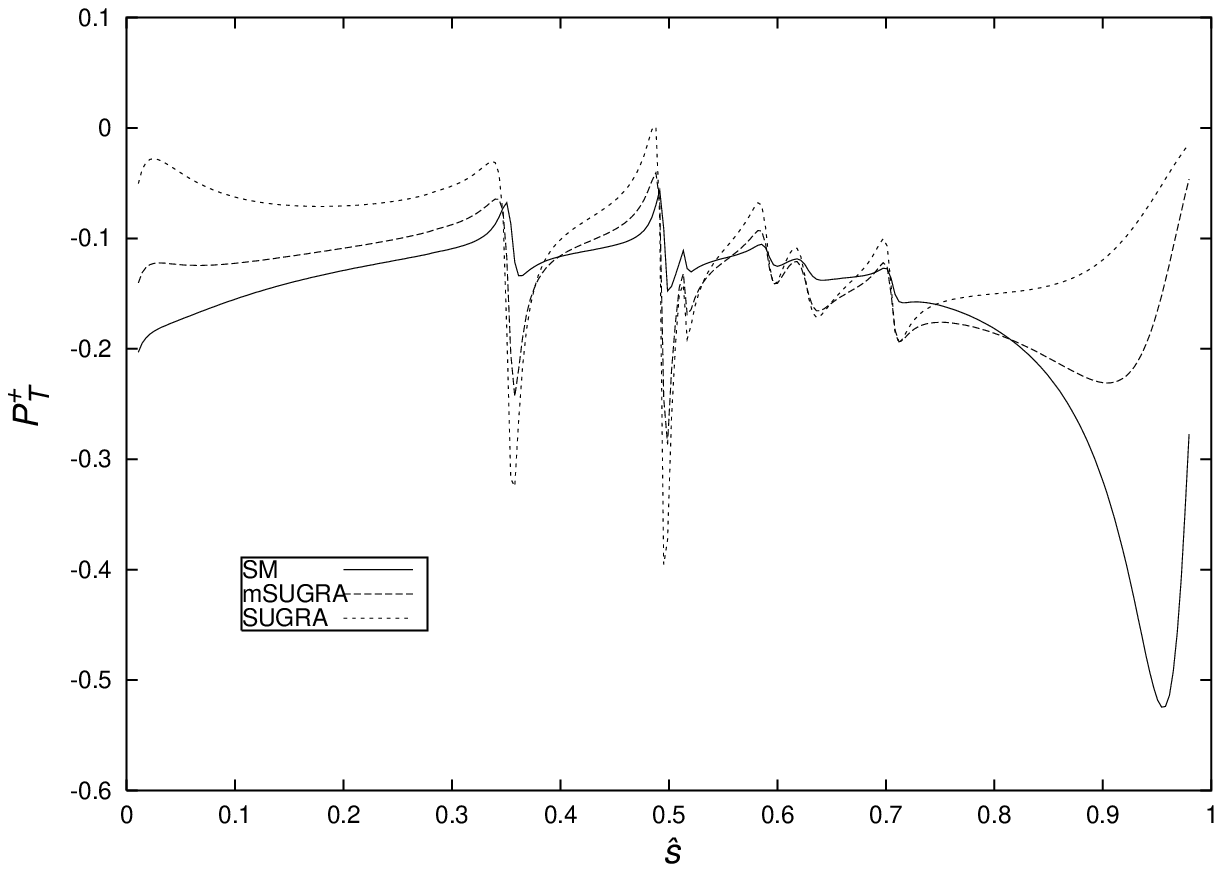,width=3.5in}
\epsfig{file=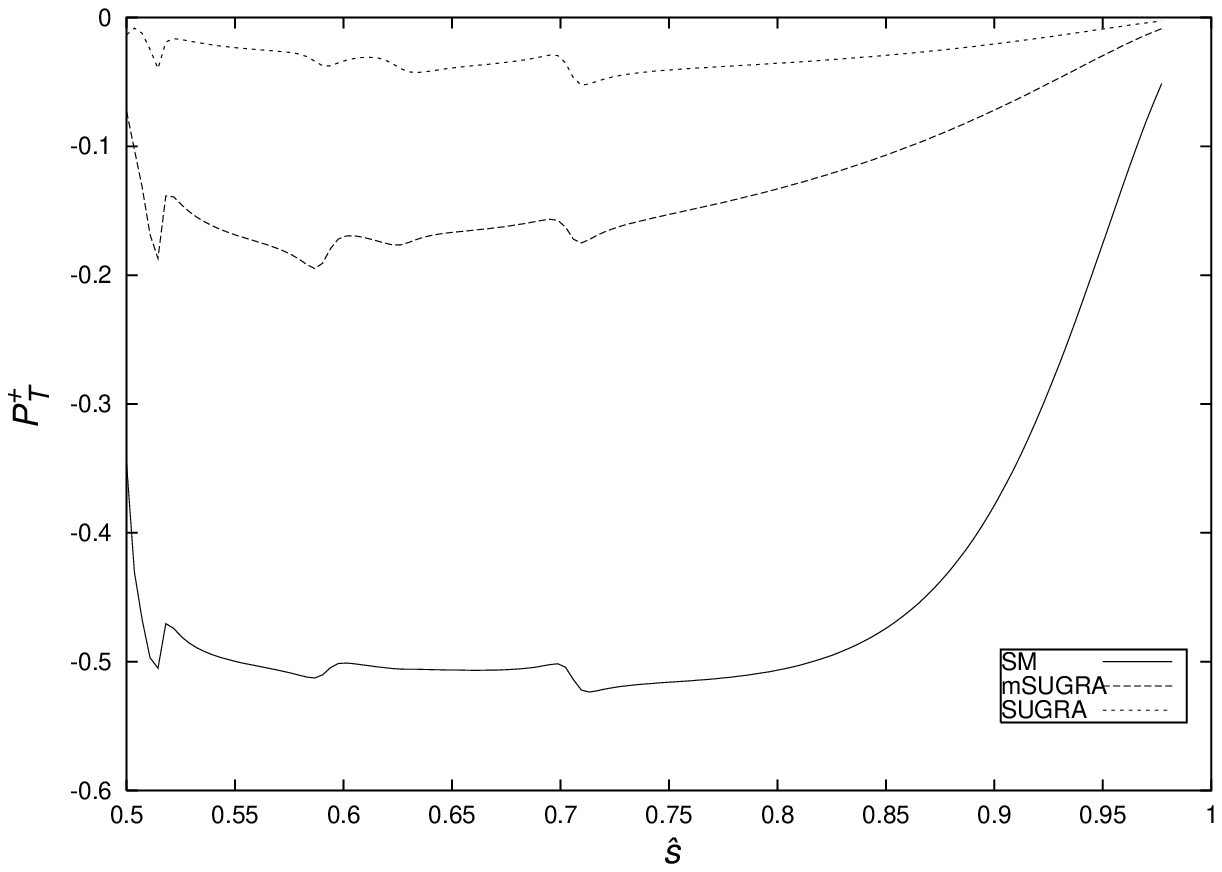,width=3.5in}
\caption{Transverse Polarization asymmetry of $\ell^+ \mu$ (left) and
$\tau$   (right). Parameters are : $m = 200 , ~M = 450 ,~ A = 0 ,
~tan\beta = 40$ and sgn($\mu$) is +ve. For SUGRA the pseudoscalar
Higgs mass is taken to be 306. All masses are in GeV} 
\label{fig:3}
\end{center}
\efig

The transverse polarization asymmetries $P_T^-$ and $P_T^+$ are :
\beqa
P_T^- &=&  \pi m_\ell
 \Bigg[ - 2 \mb2 Re(A^* B) T_1(\sh) ~+~ \frac{64 f_b^2 m_\ell}{\mb2 } 
         \ctencqtwo \drcqone T_2(\sh)    \nonumber   \\
  && +~ 8 \ctencqtwo f_b ( Re(B) T_3(\sh) ~+~ Re(C) T_4(\sh) ) 
                                         \nonumber   \\
  && +~ 8 \drcqone f_b ( Re(A) T_5(\sh) ~+~ Re(D) T_6(\sh) ) 
   \Bigg]/\bigtriangleup 
\label{sec2:eq:8}             \\ 
P_T^+ &=&  \pi m_\ell
 \Bigg[ - 2 \mb2 Re(A^* B) T_1(\sh) ~+~ \frac{64 f_b^2 m_\ell}{\mb2 } 
         \ctencqtwo \drcqone T_2(\sh)    \nonumber   \\
  && +~ 8 \ctencqtwo f_b ( Re(B) T_3(\sh) ~-~ Re(C) T_4(\sh) ) 
                                         \nonumber   \\
  && +~ 8 \drcqone f_b ( Re(A) T_5(\sh) ~-~ Re(D) T_6(\sh) ) 
   \Bigg]/\bigtriangleup
\label{sec2:eq:9}
\eeqa
with
\beqa
T_1(\sh) &=& (1 - \sh)^2 \sqrt(\sh)      \nonumber  \\
T_2(\sh) &=& \frac{(1 - 4 \mle2)}{(-1 + \sh)}  \nonumber  \\
T_3(\sh) &=& \frac{(1 - \sh) (4 \mle2 + \sh)}{(2 m_\ell + \sqrt{s})} 
                                         \nonumber  \\
T_4(\sh) &=&  ( - 2 m_\ell + \sqrt{\sh}) ( 1 + \sh)  \nonumber  \\
T_5(\sh) &=&  \frac{(4 \mle2 + \sh - 12 \mle2 \sh + \sh^2)}{(2 m_\ell
                   + \sqrt{\sh})}        \nonumber  \\
T_6(\sh) &=&  (2 m_\ell - \sqrt(\sh) ) (-1 + \sh) 
\label{sec2:eq:10}
\eeqa

\bfig[h]
\begin{center}
\epsfig{file=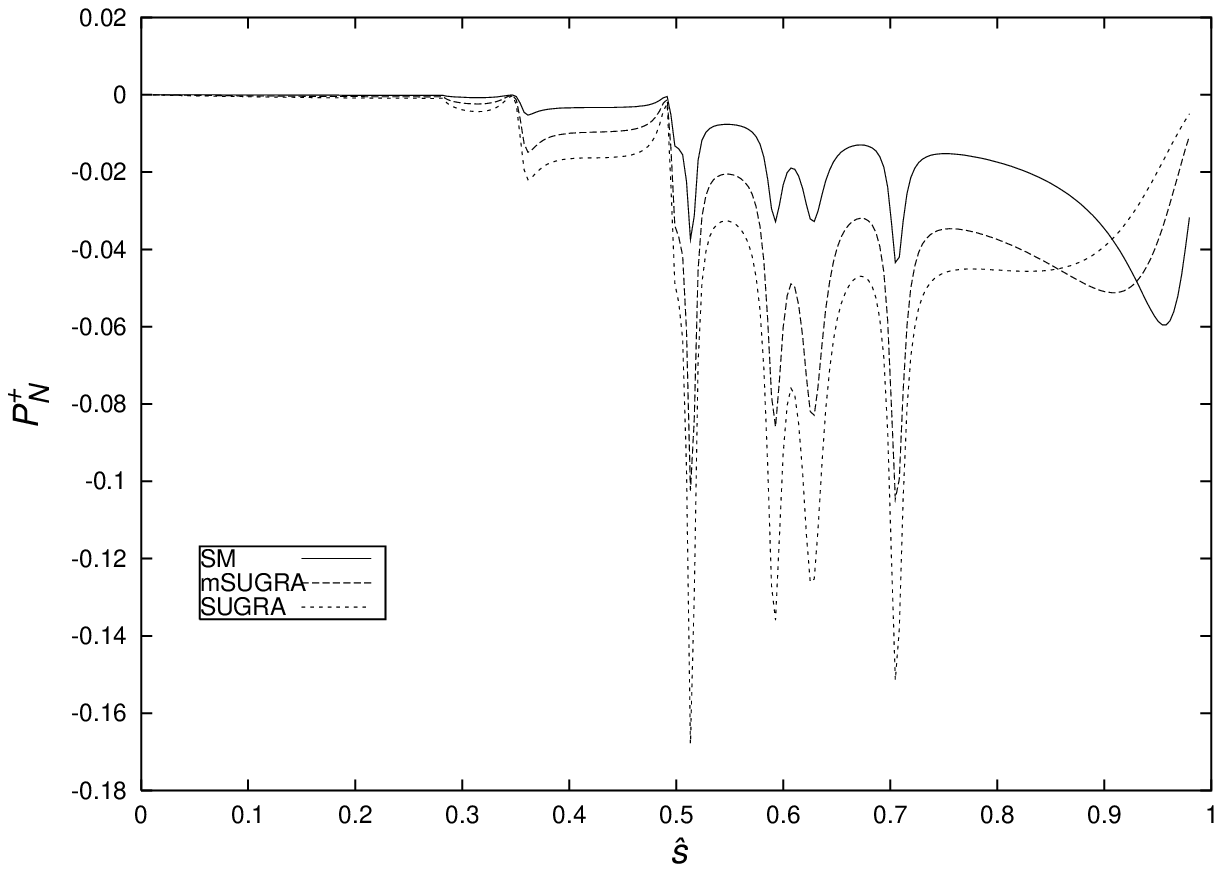,width=3.5in}
\epsfig{file=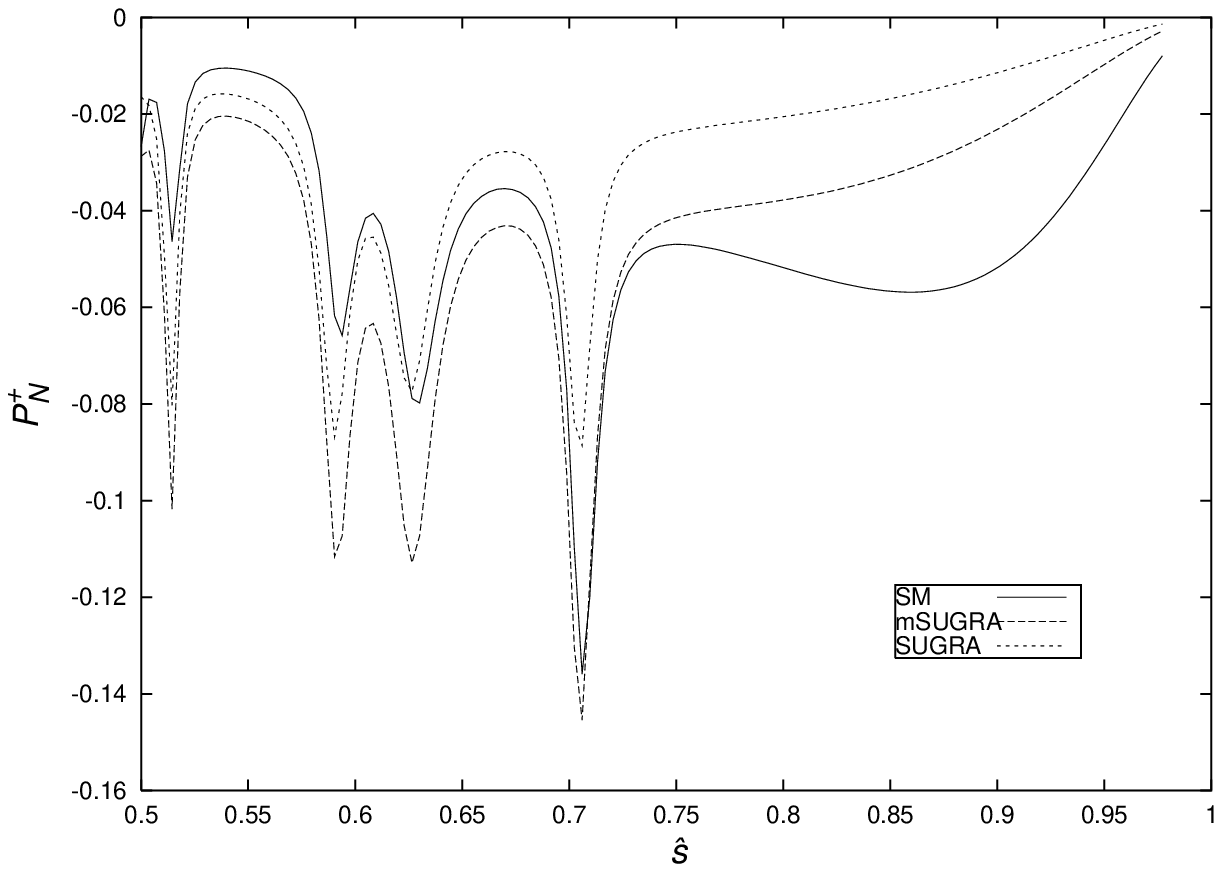,width=3.5in}
\caption{Normal Polarization asymmetry of $\ell^+ \mu$ (left) and
$\tau$   (right). Parameters are : $m = 200 , ~M = 450 ,~ A = 0 ,
~tan\beta = 40$ and sgn($\mu$) is +ve. For SUGRA the pseudoscalar
Higgs mass is taken to be 306. All masses are in GeV} 
\label{fig:4}
\end{center}
\efig

Finally the normal polarization asymmetries $P_N^-$ and $P_N^+$ are :
\beqa
P_N^- &=& m_\ell \pi 
  \Bigg[ - \mb2 Im(A^* D + B^* C) N_1(\sh)     \nonumber    \\
 &&  +~ 8 \ctencqtwo f_b \left\{ Im(A) N_2(\sh) ~+~ Im(D) N_3(\sh)
        \right\}                     \nonumber      \\
 &&  +~ 8 \drcqone f_b \left\{ Im(B) N_4(\sh) ~+~ Im(C) N_5(\sh)
         \right\} 
  \Bigg]                \\
\label{sec2:eq:11}
P_N^+ &=& m_\ell \pi 
  \Bigg[  \mb2 Im(A^* D + B^* C) N_1(\sh)     \nonumber    \\
 &&  +~ 8 \ctencqtwo f_b \left\{ Im(A) N_2(\sh) ~-~ Im(D) N_3(\sh)
        \right\}                     \nonumber      \\
 &&  +~ 8 \drcqone f_b \left\{ Im(B) N_4(\sh) ~-~ Im(C) N_5(\sh)
         \right\} 
  \Bigg] 
\label{sec2:eq:12}
\eeqa
with
\beqa
N_1(\sh) &=&  ( - 1 + \sh ) \sqrt(\sh) \faco    \nonumber  \\
N_2(\sh) &=& \frac{\sh ( 1 + \sh ) \faco}{(2 m_\ell + \sqrt{\sh})} 
                                     \nonumber    \\
N_3(\sh) &=&  \frac{( 1 - \sh ) \sh \faco}{(2 m_\ell + \sqrt{\sh})} 
                                     \nonumber    \\
N_4(\sh) &=&  \frac{(1 - \sh) \sh \faco }{(2 m_\ell + \sqrt{\sh})} 
                                     \nonumber    \\
N_5(\sh) &=&  \frac{ \sh ( 1 - 8 \mle2 + \sh ) \faco }{(2 m_\ell +
              \sqrt{\sh})}          
\label{sec2:eq:13}
\eeqa

\section{\label{section:4} Combined analysis of the polarization
asymmetries} 

One can get useful information about new physics by observing the
asymmetries of $\ell^-$ and $\ell^+$ and the combination of these
asymmetries. As has been shown in many other earlier works (in model
independent ways) regarding these asymmetries in decay modes $B \to
X_s \ell^- \ell^+$ and $B \to K^* \ell^- \ell^+$ \cite{Fukae:1999ww}
that within SM some linear combination of these asymmetries
vanish. The major results of those investigations were (within 
Standard Model) 
: 
\begin{itemize}
\item{} $P_L^- ~+~ P_L^+ ~=~ 0$.
\item{} $P_T^- ~-~ P_T^+ ~\approx~ 0$.
\item{} $P_N^- ~+~ P_N^+ ~=~ 0 $.
\end{itemize}
It was argued that any deviations from above result will give a
definite signal for new physics.  

\par We will analyse the same combination of the various polarization
asymmetries within the radiative dileptonic decay (\bsllg) and will
try to figure out what happens to these combinations within SM and if
SUSY effects are included. 

\bfig[h]
\begin{center}
\epsfig{file=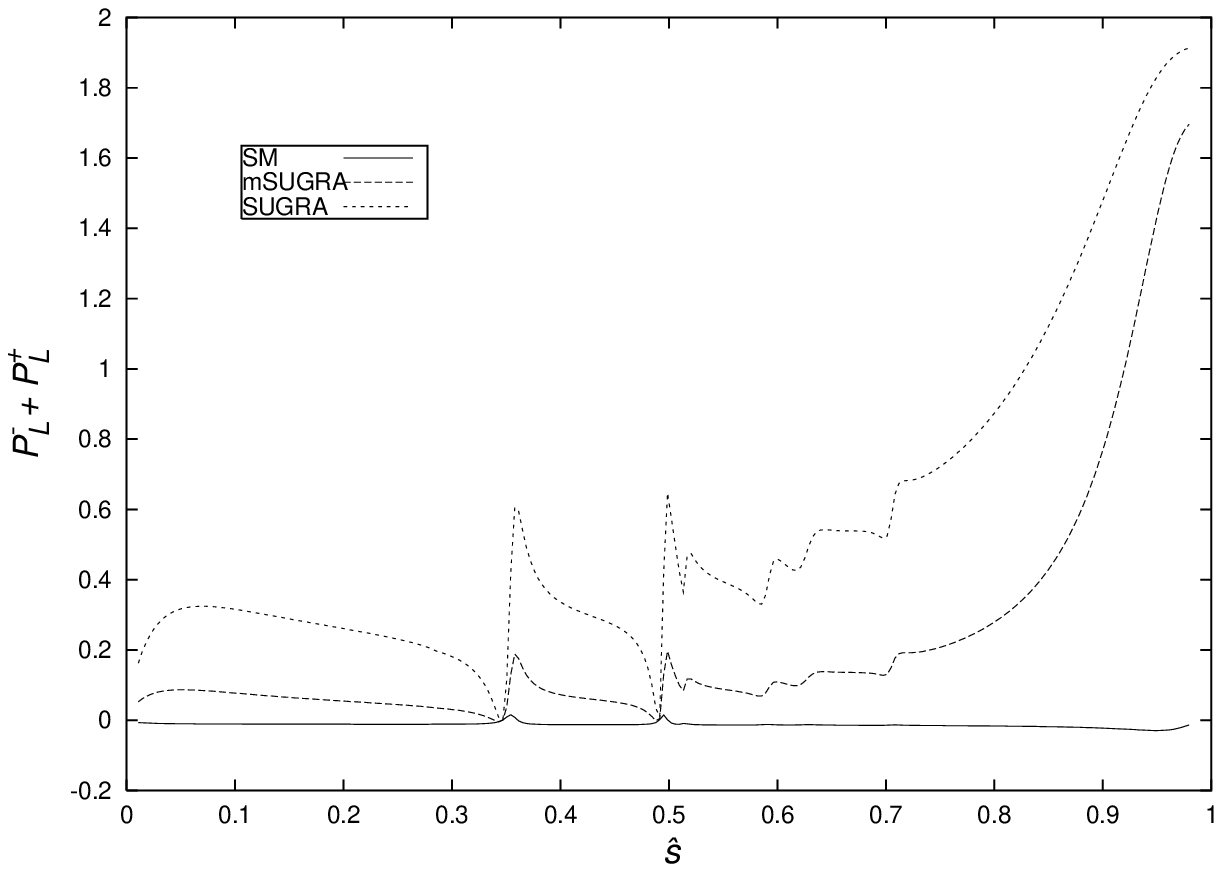,width=3.5in}
\epsfig{file=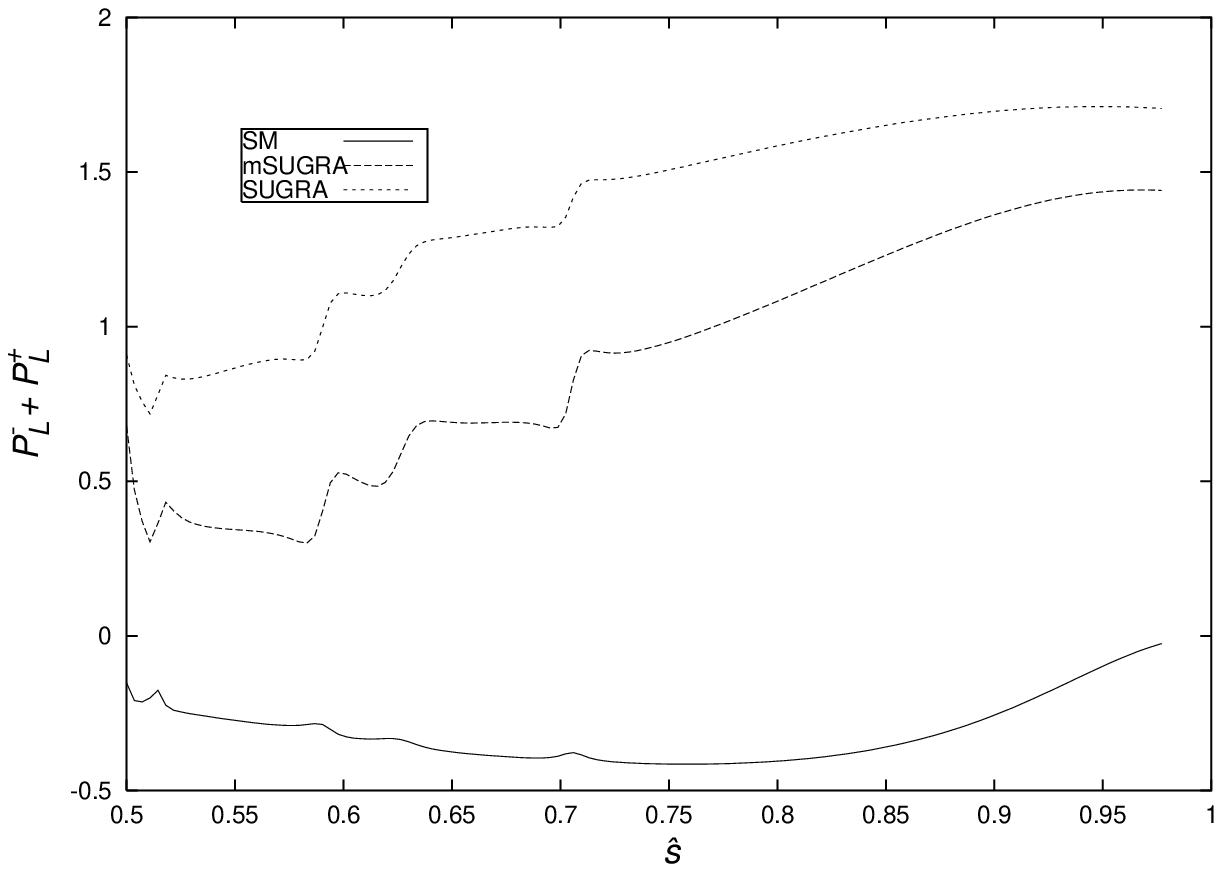,width=3.5in}
\caption{Sum of Longitudinal Polarization asymmetries of $\ell^-$ and
$\ell^+$, $\mu$ (left) and $\tau$   (right). Parameters are : $m = 200
, ~M = 450 ,~ A = 0 , ~tan\beta = 40$ and sgn($\mu$) is +ve. For SUGRA
the pseudoscalar Higgs mass is taken to be 306. All masses are in GeV}
\label{fig:5}
\end{center}
\efig

{\bf (A)} For $P_L^- + P_L^+$ the result is :
\beqa
P_L^- + P_L^+ 
 &=& 64 ~ f_b ~ \mle2 ~
  \Bigg[ 4 \frac{f_b}{\mb2 } \ctencqtwo \drcqone L_2(\sh)  \nonumber \\
 &&  +~ \ctencqtwo  Re(B) L_3(\sh) ~+~ \drcqone Re(A) L_5(\sh) 
  \Bigg]/\bigtriangleup
\label{sec3:eq:1}
\eeqa
we can very easily see that within SM (when $\cq1 $ and $\cq2 $ are
zero) the sum of longitudinal polarization asymmetries of $\ell^-$ and
$\ell^+$ doesn't vanish. 

\bfig[ht]
\begin{center}
\epsfig{file=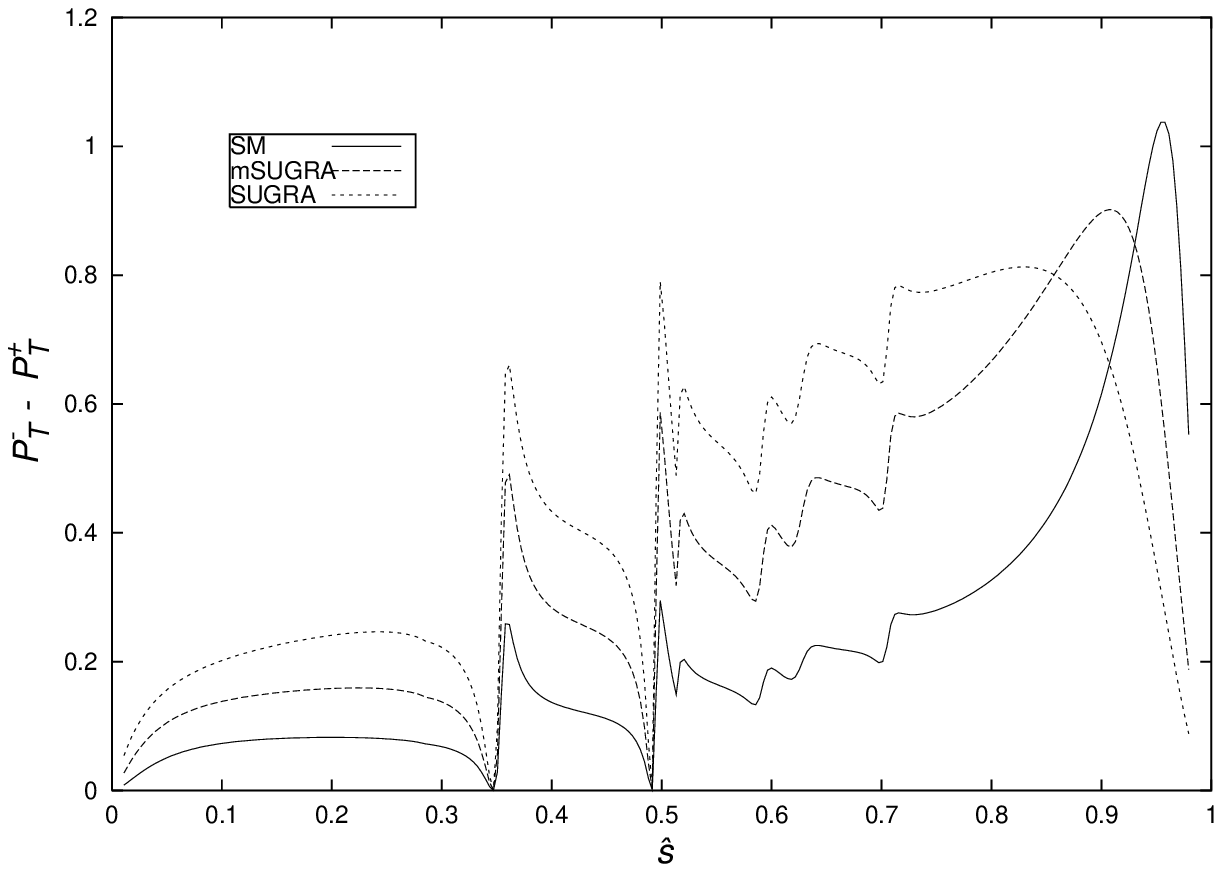,width=3.5in}
\epsfig{file=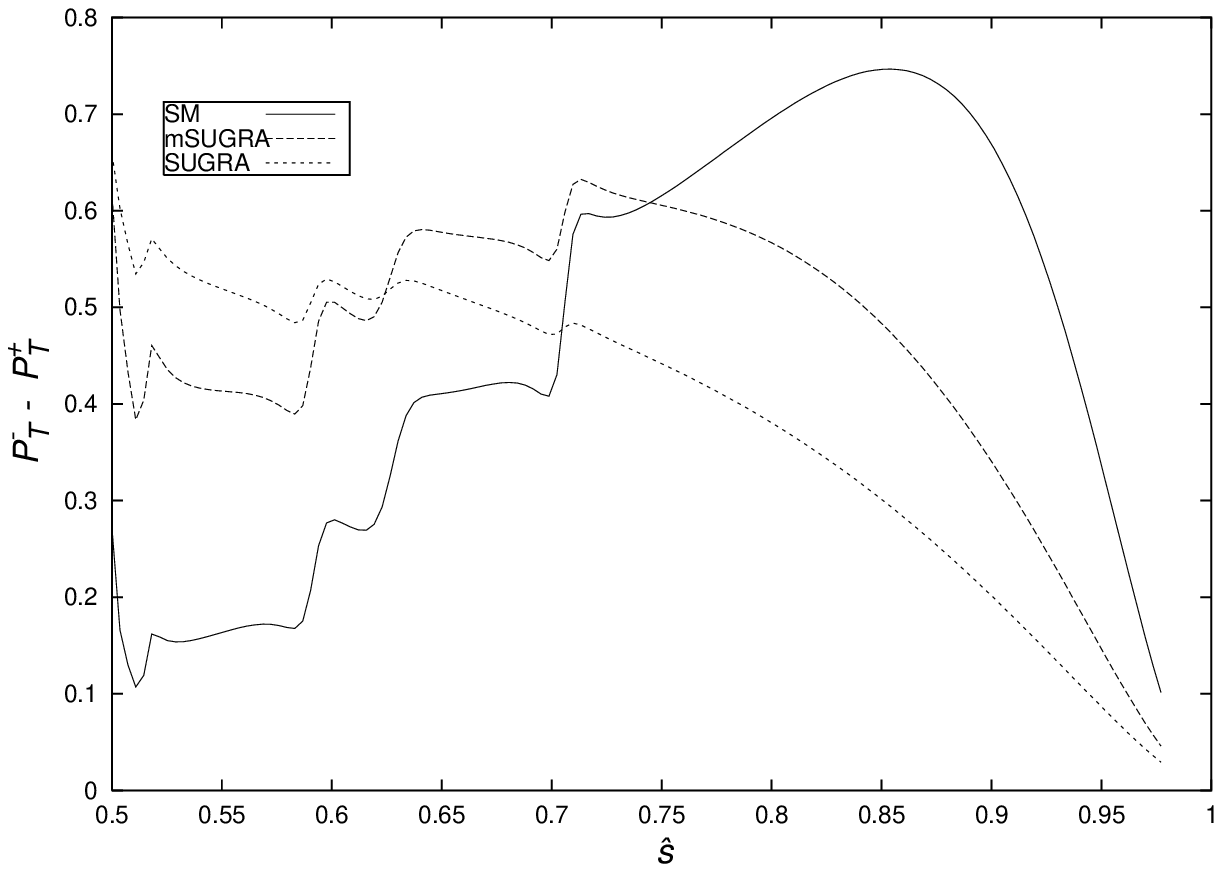,width=3.5in}
\caption{Difference of Transverse Polarization asymmetries of $\ell^-$
and $\ell^+$, $\mu$ (left) and $\tau$   (right). Parameters are : $m =
200 , ~M = 450 ,~ A = 0 , ~tan\beta = 40$ and sgn($\mu$) is +ve. For
SUGRA the pseudoscalar Higgs mass is taken to be 306. All masses are
in GeV}  
\label{fig:6}
\end{center}
\efig

{\bf (B)} For $P_T^- - P_T^+$ the result is :
\beq
P_T^- - P_T^+ ~=~ 16 \pi m_\ell f_b
  \Bigg[ \ctencqtwo Re(C) T_4(\sh) ~+~ \drcqone Re(D) T_6(\sh)
  \Bigg]/\bigtriangleup
\label{sec3:eq:2}
\eeq
we can again see the same pattern that within SM the difference of the
transverse polarization asymmetries of $\ell^-$ and $\ell^+$ doesn't
vanish. 

\bfig[ht]
\begin{center}
\epsfig{file=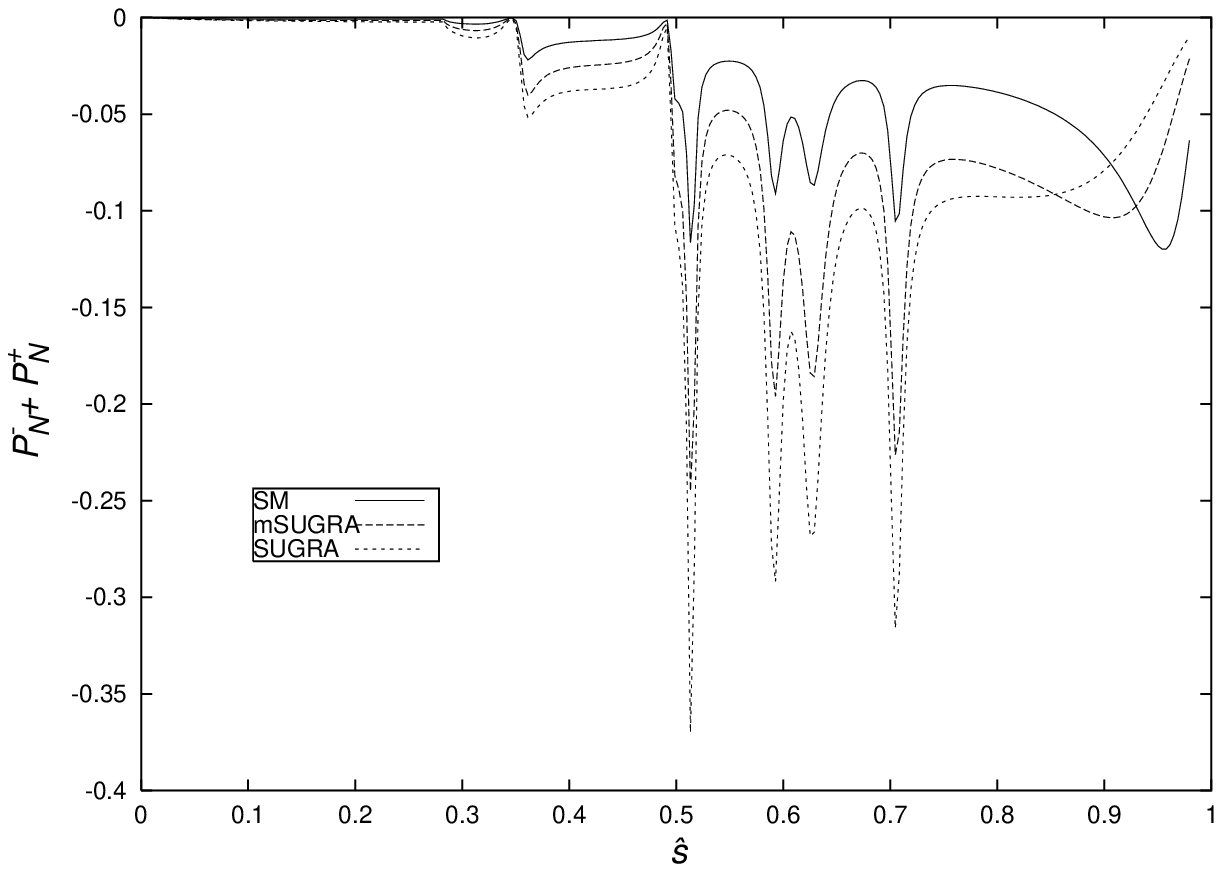,width=3.5in}
\epsfig{file=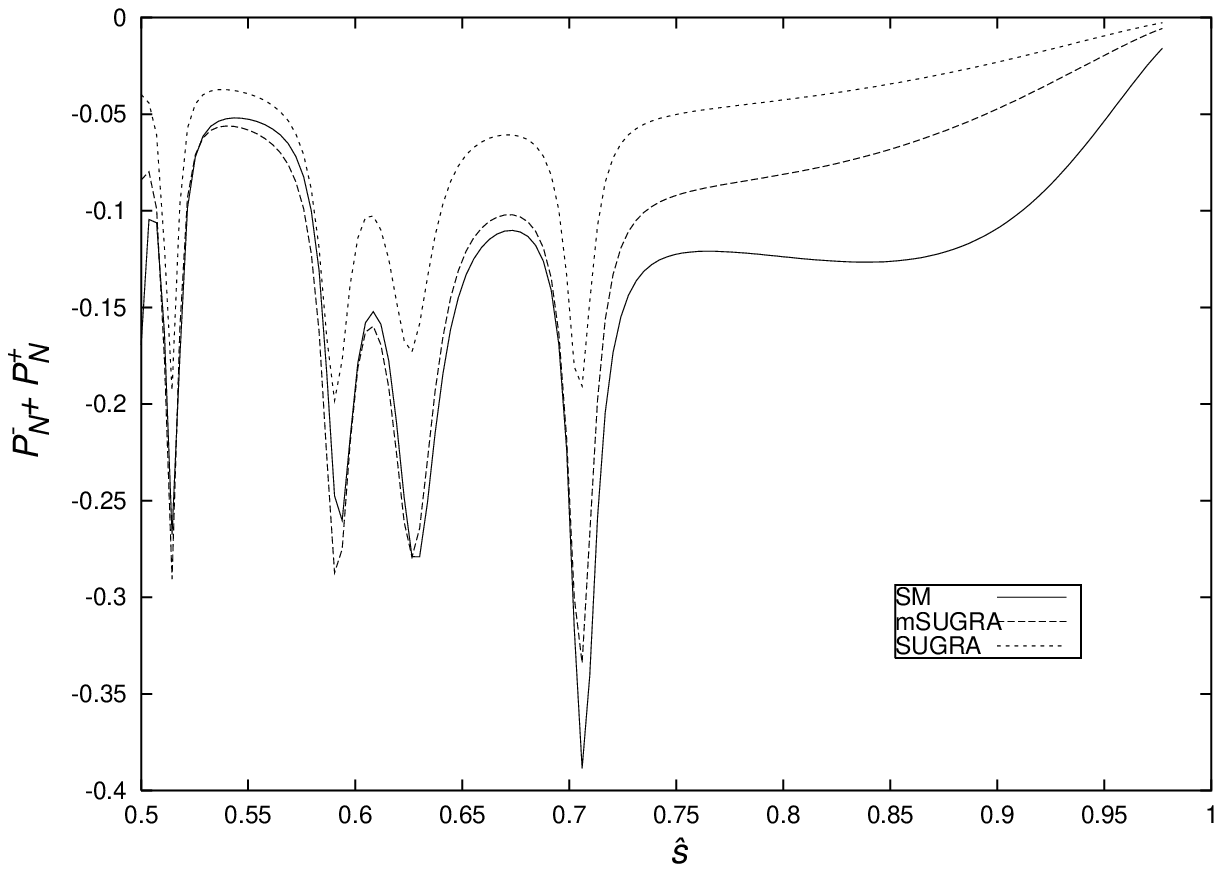,width=3.5in}
\caption{Sum of Normal Polarization asymmetries of $\ell^-$ and
$\ell^+$, $\mu$ (left) and $\tau$   (right). Parameters are : $m = 200
, ~M = 450 ,~ A = 0 , ~tan\beta = 40$ and sgn($\mu$) is +ve. For SUGRA
the pseudoscalar Higgs mass is taken to be 306. All masses are in GeV}
\label{fig:7}
\end{center}
\efig

{\bf (C)} For $P_N^- + P_N^+$ the result is :
\beq
P_N^- + P_N^+ ~=~ 16 \pi m_\ell 
  \Bigg[ \ctencqtwo Im(A) N_2(\sh) ~+~ \drcqone Im(B) N_4(\sh)
  \Bigg]
\label{sec3:eq:3}
\eeq
we have the repetition of the same pattern that the sum of the
polarization asymmetries of $\ell^-$ and $\ell^+$ doesn't vanish. 

\section{\label{section:4b} Photon polarization asymmetry}
In radiative dileptonic decay mode of B-meson (\bsllg ) even the final
state photon can emerge with a definite polarization. So one can study
the effects of polarized photon also in this particular decay
mode. Here we introduce such a variable which we call {\sl photon
polarization asymmetry}. Defined as :

\beq
H ~=~ \frac{ \frac{d \Gamma (\epsilon^* = \epsilon_1)}{d \sh} 
      - \frac{d \Gamma (\epsilon^* = \epsilon_2)}{d \sh} }
      { \frac{d \Gamma (\epsilon^* = \epsilon_1)}{d \sh} 
      + \frac{d \Gamma (\epsilon^* = \epsilon_2)}{d \sh} }
\label{sec4:eq:1}
\eeq
where $\epsilon_1$ and $\epsilon_2$ are the two polarization states
which a photon can have \footnote{photon being mass-less particle can
only have two polarization states which we have called $\epsilon_1$ and
$\epsilon_2$ which actually are conjugate to each other. These states
can equally be called positive and negative helicity states
respectively}. 

\par Now in order to evaluate the variable $H$ we have to consider
polarized photon in decay rate calculation. In CM (center of mass
frame of dileptons) the various four vectors ( of $B_s$ meson, photon
, leptons and polarizations of photons) can be taken as \footnote{here
we are choosing leptons to be lying in YZ-plane and B-meson is moving
along Z-direction} :
\beqa
P_B &=&  (~E_B ~,~ 0~ ,~ 0 ~,~ p_B ~)  \nonumber   \\
q   &=&  ( ~p_B ~,~ 0 ~,~ 0 ~,~ p_B ~)    \nonumber   \\
p_- &=&  \left( ~\frac{\sqrt{s}}{2} ~,~ 0 ~,~ p~ Sin\theta ~,~ - ~p~
Cos\theta ~ \right) \nonumber \\
p_+ &=&  \left( ~ \frac{\sqrt{s}}{2} ~,~ 0 ~,~ -~ p~ Sin\theta ~,~  p~
Cos\theta ~\right) \nonumber   \\
\epsilon_1 &=&  {1 \over \sqrt{2}} ~(~0 ~,~ 1 ~,~ i ~,~ 0 ~)
\nonumber  \\ 
\epsilon_2 &=&  {1 \over \sqrt{2}} ~(~0 ~,~ 1 ~,~ - ~ i ~,~ 0 ~) 
\label{sec4:eq:2}
\eeqa
where $p_B ~=~ \frac{(m_{B_s}^2 - s)}{2 \sqrt{s}}$ and $p ~=~
\frac{\sqrt{s}}{2} ~ \sqrt{1 - \frac{4 m_\ell^2}{s}}$. 

\par Using the above identities one can easily calculate $H$ as :
\beqa
H &=& \Bigg[ \frac{8 \mb2 }{3} Re(A^* B) (1 - \sh )^2 (2 \mle2 + \sh )
     ~+~  \frac{8 \mb2 }{3} Re(C^* D) (1 - \sh )^2 ( - 4 \mle2 + \sh )
                      \nonumber     \\
  && +~  32 f_b \mle2 \ctencqtwo Re(B) \frac{(1 - \sh ) ln(\zh
           )}{\faco}       \nonumber   \\
  &&  +~ 32 f_b \mle2 \drcqone  Re(A) 
       \frac{ \sh ( 8 \mle2 - 2 \sh - \faco ln(\zh) ( - 1 + 4 \mle2 +
           \sh ))}{4 \mle2 - \sh} 
      \Bigg]/\bigtriangleup
\label{sec4:eq:3}
\eeqa

\bfig[ht]
\begin{center}
\epsfig{file=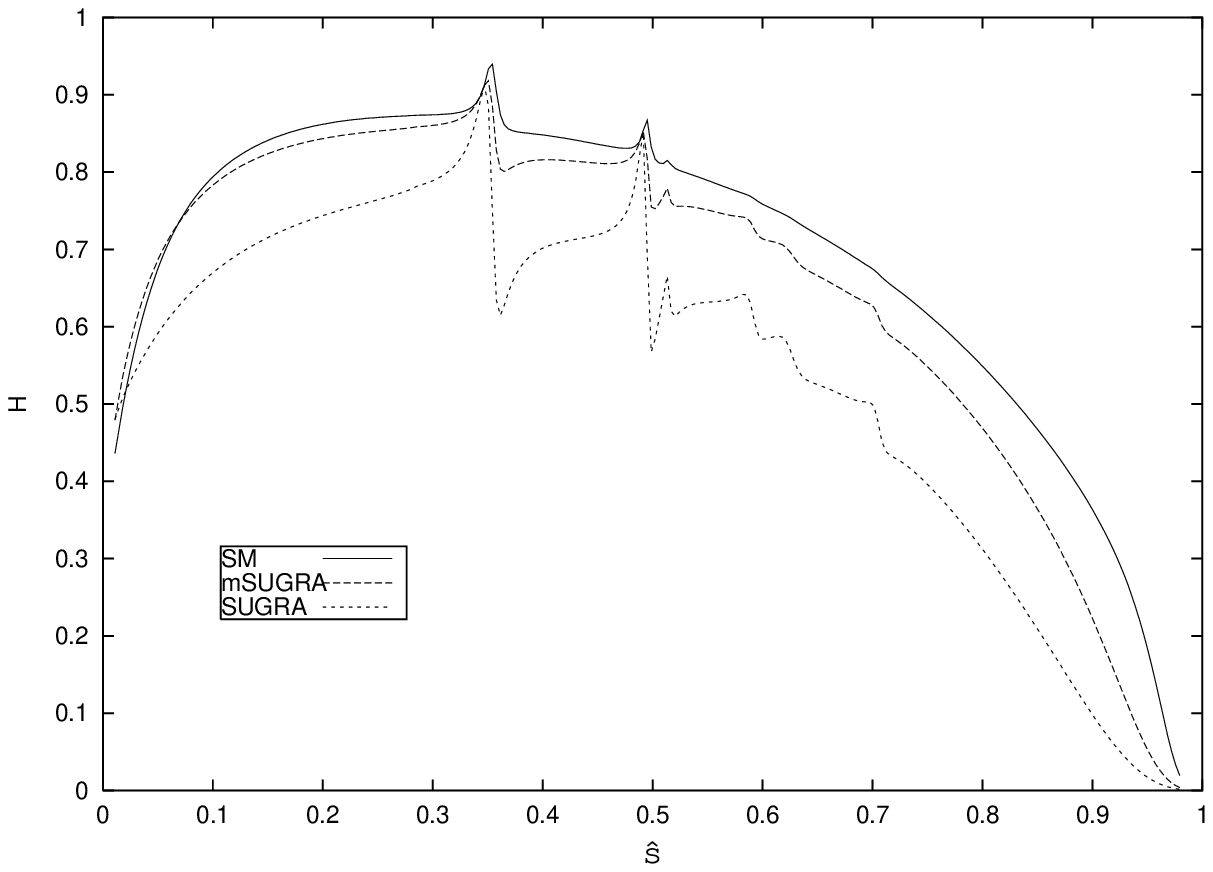,width=3.5in}
\epsfig{file=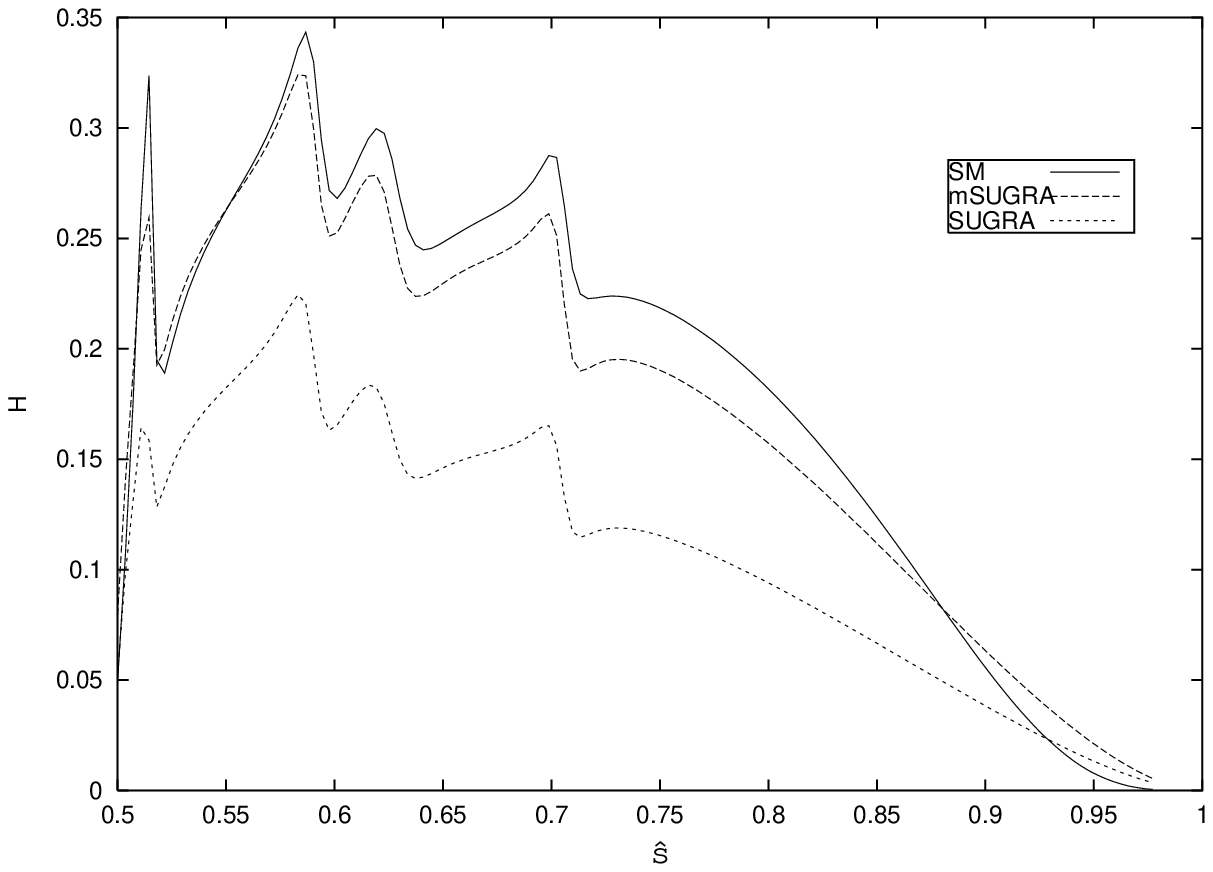,width=3.5in}
\caption{Photon Polarization asymmetry , $\mu$ (left) and $\tau$
(right). Parameters are : $m = 200 , ~M = 450 ,~ A = 0 , ~tan\beta =
40$ and sgn($\mu$) is +ve. For SUGRA the pseudoscalar Higgs mass is
taken to be 306. All masses are in GeV} 
\label{fig:8}
\end{center}
\efig

\section{\label{section:5} Results and Discussion}

We have performed the numerical analysis of the forward-backward
asymmetry and various polarization asymmetries whose analytical
expressions are given in earlier sections. 

\par We are working under MSSM, which is the simplest having the least
number of parameters among the various extensions of the SM. But even
the MSSM has a very large number of parameters making it very
difficult to do any reasonable phenomenology in such a large parameter
space. We therefore focus on some unified models which reduce this
large number of parameters to a manageable number. One of the most
favorite such model is the Supergravity (SUGRA) model. In SUGRA model
in addition to the SM parameters we have : $m_0$ (unified mass of all
the scalars), M (unified mass of all the gauginos), A (unified
trilinear couplings), $tan\beta$ (ratio of the VEV of the two Higgs)
and $sgn(\mu)$ as parameters \footnote{convention about sgn($\mu$)
which we are following is such that $\mu$ occurs in chargino mass
matrix with positive sign} . This sort of model is called Minimal
Sugra (mSUGRA) model. 

\par It has been well emphasized in many works
\cite{Choudhury:1999ze,RaiChoudhury:1999qb,goto1} that it is not 
necessary that all the scalars have a unified mass at GUT scale. To
suppress $K^0-{\bar K}^0$ mixing its sufficient to have the universal
mass to all the squarks. We will hence explore this type of SUGRA
model also where the squarks and Higgs sector has different universal 
mass. For the results shown in Figures all the MSSM parameters were
evolved from GUT scale to electroweak scale using SuSpect
\cite{djouadi}. 
For MSSM parameter space analysis we have taken a 95\% CL bound
of \cite{expbsg} :
$$ 2 \times 10^{-4} ~<~ Br(B \to X_s \gamma ) ~<~ 4.5 \times 10^{-4}
$$ 
which is in agreement with CLEO and ALEPH results. 

In the SUGRA model where the condition of universality of the scalar
masses has been relaxed we take the mass of pseudo-scalar Higgs to be
a parameter. In Figure (\ref{fig:1}) we have plotted the 
forward-backward asymmetry of the lepton in SM, mSUGRA and SUGRA
model. As we can see that both SUGRA and mSUGRA gives fairly large
deviation from the SM values specially when we have $\tau$ in the
final states. In Figures (\ref{fig:2} , \ref{fig:3} , \ref{fig:4}) we
have plotted the longitudinal, transverse and normal polarization
asymmetries respectively of $\ell^+$. As we can see that all the three
asymmetries shows fairly large deviation from their respective SM
values. The deviation from the SM values is largely because of the new
Wilson coefficients $\cq1 $ and $\cq2 $ . These coefficients depends
crucially on the two MSSM parameters the Higgs mass and $tan\beta$. So
for relatively large $tan\beta$ (we have taken this to be 40 here in
our calculations) we can have very large variations in the
polarization asymmetries. 

\par Another important observation here is that in earlier model
independent analysis of the B-meson decay modes like $B \to X_s \ell^+
\ell^-$ and $B \to K^* \ell^+ \ell^-$ \cite{Fukae:1999ww} the sum of
longitudinal and normal polarization asymmetries (of $\ell^-$ and
$\ell^+$) vanishes separately in SM . Also the difference of the
transverse polarization asymmetries of $\ell^-$ and $\ell^+$
vanishes. But here in the radiative dileptonic decay mode (\btosllg)
this hasn't happened. So that means that the results quoted in Fukawe
\etal and Aliev \etal \cite{Fukae:1999ww} (that these quantities
identically vanish in SM) was a process dependent statement. Here in
the process we are considering the $C_{10}$ (which is non-zero in SM)
is behaving exactly like $\cq2 $. As we can see from figures
(\ref{fig:5} ,\ref{fig:7}) that the sum of polarization
asymmetries , of $\ell^+$ and $\ell^-$ (for longitudinal and normal)
doesn't vanish even within SM. From figure(\ref{fig:6}) we can
conclude that the difference of the transverse asymmetries also
doesn't vanish within SM . Figure(\ref{fig:8}) shows the photon
polarization asymmetry for completenes. This is much more challenging
experimentally to measure as compared to the $\tau^\pm$ polarization
asymmetries. So the lepton polarization asymmetries (as compared to
photon one) are the best places to look for physics beyond SM in FCNC
semi-leptonic B-decays. 

\par From our theoretical and numerical analysis we can thus conclude
:
\begin{enumerate}
\item{} From the analysis of the FB-asymmetry, all the three
polarization asymmetries associated with the final state leptons we
can say that the deviation in these quantities from their respective
SM values is fairly substantial almost over the whole region of
dileptons invariant mass. 
\item{} As noted in earlier papers regarding general polarization
asymmetries in various semi-leptonic decay modes of B-meson
\cite{Fukae:1999ww} that the sum of the longitudinal and normal
polarization asymmetries (independently) for lepton ($\ell^-$) and
anti-lepton ($\ell^+$) vanishes in SM . Similarly the difference of 
transverse polarization asymmetries of lepton ($\ell^-$) and
anti-lepton ($\ell^+$) also vanishes in SM. But here in radiative
dileptonic decay mode (\bsllg) this is no-longer true. Here the sum of
the longitudinal polarization asymmetries of $\ell^-$ and $\ell^+$ in
SM is very small (not exactly zero) for leptons to be muon (see
fig.(\ref{fig:5})) and the deviation is fairly large in mSUGRA and
SUGRA model. 
\item{} The {\sl photon polarization asymmetry} also shows a large
variation from the respective SM value both for muons and tau but this
measurement would be much more difficult experimentally. Nonetheless
this is still another kinematical variable which could be used to study
radiative decays. 
\item{} As we can see that the variation of all the kinematical
variables which we have analysed in our work shows large deviation
from the respective SM values. But the deviation from the SM is much
more enhanced in SUGRA model then in mSUGRA model. The reason for this
is the structure of the new Wilson coefficients $\cq1 $ and $\cq2
$. These coefficients depends on Higgs mass (in fact the dependence of
these coefficients to Higgs mass is inverse). So as the Higgs mass
increases the value of these coefficients decreases. In mSUGRA
framework where all the scalars have a common unified mass the Higgs
mass turn out to be very high ( and hence value of these coefficients
small). Whereas in SUGRA framework as the Higgs sector has a different
unification so the masses could be low ( and hence fairly large values
of the new coefficients). 
\end{enumerate}

So in brief one can say that even the radiative decay mode can be a
useful probe in finding out the SUSY signatures and can also probe
into the structure of effective Hamiltonian. 



\end{document}